\documentclass[sn-mathphys,Numbered]{sn-jnl}

\usepackage{graphicx}%
\usepackage{multirow}%
\usepackage{amsmath,amssymb,amsfonts}%
\usepackage{amsthm}%
\usepackage{mathrsfs}%
\usepackage[title]{appendix}%
\usepackage{xcolor}%
\usepackage{textcomp}%
\usepackage{manyfoot}%
\usepackage{booktabs}%
\usepackage{algorithm}%
\usepackage{algorithmicx}%
\usepackage{algpseudocode}%
\usepackage{listings}%

\usepackage{mathtools}
\usepackage{braket}
\usepackage{tikz}
\usetikzlibrary{patterns}
\usepackage{pgfplots}
\pgfplotsset{compat=1.17}
\usepackage{subcaption}
\usepackage{color}
\usepackage{bm}
\usepackage{cancel}
\usepackage{tensor}
\usepackage{stackrel}
\usepackage{tikz-cd}
\usepackage{tabularx}

\begin{document}

\title{Gravitational-wave matched filtering with variational quantum algorithms}

\author*[1,2]{\fnm{Jason} \sur{Pye}}\email{jason.pye@su.se}
\author[1]{\fnm{Edric} \sur{Matwiejew}}
\author[1]{\fnm{Aidan} \sur{Smith}}
\author[1,3]{\fnm{Manoj} \sur{Kovalam}}
\author[1]{\fnm{Jingbo B.} \sur{Wang}}
\author[1,3]{\fnm{Linqing} \sur{Wen}}

\affil[1]{\orgdiv{Department of Physics}, \orgname{The University of Western Australia}, \orgaddress{35 Stirling Hwy, \city{Crawley}, \state{WA} \postcode{6009}, \country{Australia}}}

\affil[2]{\orgname{Nordita, Stockholm University and KTH Royal Institute of Technology},\\ \orgaddress{Hannes Alfv\'ens v\"ag 12, SE-106 91 \city{Stockholm}, \country{Sweden}}}

\affil[3]{\orgname{Australian Research Council Centre of Excellence for Gravitational Wave Discovery (OzGrav)}, \orgaddress{35 Stirling Hwy, \city{Crawley}, \state{WA} \postcode{6009}, \country{Australia}}}

\abstract{
In this paper, we explore the application of variational quantum algorithms designed for classical optimization to the problem of matched filtering in the detection of gravitational waves. Matched filtering for detecting gravitational wave signals requires searching through a large number of template waveforms, to find one which is highly correlated with segments of detector data. This computationally intensive task needs to be done quickly for low latency searches in order to aid with follow-up multi-messenger observations. The variational quantum algorithms we study for this task consist of quantum walk-based generalizations of the Quantum Approximate Optimization Algorithm (QAOA). We present results of classical numerical simulations of these quantum algorithms using open science data from LIGO. These results show that the tested variational quantum algorithms are outperformed by an unstructured restricted-depth Grover search algorithm, suggesting that the latter is optimal for this computational task.
}


\maketitle


\section{Introduction}

Gravitational wave (GW) astronomy has seen many successes since the discovery of the first binary black hole merger GW150914 by the advanced Laser Interferometer Gravitational-Wave Observatory (LIGO) detectors in 2015 \cite{AbbottEtal2016}.
So far, the GW detectors have completed three observing runs, publishing a total of 93 GW detections.

The search for GW signals in detector data is performed by search pipelines in two timescales: low latency and offline.
The low latency pipelines are built to facilitate rapid detection of GWs, which allows for follow-up multi-messenger observations, while the offline pipelines are used to determine the accuracy and significance of the events, which are published through the GW transient catalog \cite{AbbottEtal2021,AbbottEtal2023}.
Most of these searches employ matched filtering to detect compact binary coalescences (CBCs).
Matched filtering is the optimal (linear) signal processing method used to detect known signals within Gaussian noise \cite{AllenEtal2012}.
It involves correlating a known CBC waveform, called a template, with detector data to find a match.
In the presence of a signal, the matched filter maximizes the signal-to-noise ratio.

Matched filtering is an easy step if the signal waveform is known.
However, because the shape of the waveform can vary significantly depending on physical properties of the CBC, detecting an unknown CBC signal within noise requires performing a blind search through many different templates (scaling linearly with the number of templates).
Typically, the GW search pipelines employ $\sim 10^5$--$10^6$ templates to perform matched filtering, which is a computationally intensive task.
Further, with the continuing advances made to the detectors and their detection sensitivities, the range of physical parameters (producing different waveforms) covered by the detectors is significantly increased.
Including this increased volume of parameter space in the search is expected to cause the number of required templates to grow by another order of magnitude \cite{KrastevEtal2021,HuertaEtal2019,HarryEtal2016,HuertaEtal2017}.
Therefore, keeping pace with future scientific demand will require searching through ever larger numbers of templates, which puts a significant strain on computational resources, especially in the case of low latency searches which require speed.

Here we investigate the potential of quantum algorithms for this computational task.
The rapid progress in the development of quantum computers has spurred much interest in finding applications where they demonstrate advantages over their classical counterparts, particularly for problems which are intractable for classical computers.
However, current quantum devices are not error-corrected and thus produce noisy outputs.
This has resulted in approaches to design quantum algorithms which are resistant to the errors occurring in this so-called noisy intermediate-scale quantum (NISQ) era \cite{Preskill2018}.
One such approach is to use variational quantum algorithms (VQAs), which comprise a set of unitary operations which depend on classical variational parameters \cite{CerezoEtal2021}.
These parameters are optimized using successive measurements of the output of the quantum circuit, in order to drive the system toward the desired output.
There has been significant effort in developing VQAs for classical and quantum optimization \cite{FarhiGoldstoneGutmann2014,PeruzzoEtal2014,Preskill2018,CerezoEtal2021}.

The challenge of searching through GW templates for matched filtering seems a suitable candidate application for quantum computing.
Indeed, Grover's search algorithm is one of the fundamental algorithms in quantum computing \cite{Grover1996}.
In \cite{GaoEtal2022}, it was shown how Grover's algorithm can be used to obtain a quadratic speedup in the number of templates for GW matched filtering.
The number of qubits required to perform this algorithm was improved upon in \cite{MiyamotoEtal2022}, while maintaining the quadratic speedup.
Other work on using quantum computing for GW astronomy includes \cite{EscrigEtal2023}, which proposes a quantum algorithm for GW parameter estimation, and \cite{HayesEtal2023}, which demonstrates an algorithm for amplitude-encoding of GW templates for potential future use in quantum machine learning models.

However, a study of the performance of VQAs for GW matched filtering has not previously been done.
Although Grover's algorithm can provide a speedup for this problem \cite{GaoEtal2022}, it is not considered to be a near-term algorithm.
Furthermore, even though it is well-known that Grover's algorithm is optimal for a general unstructured search problem \cite{Zalka1999}, it is possible that one can improve upon it by making use of structure present in particular classes of problems \cite{AaronsonAmbainis2009,McCleanEtal2021}.
Thus, even for fully error-corrected quantum computers, one should also consider structured quantum approaches to GW matched filtering.

Here we will examine the performance, through numerical simulation, of various VQAs for a particular GW template search problem, and compare their performance to Grover's search algorithm (as proposed in \cite{GaoEtal2022}).
In particular, we will formulate the GW template search as an optimization problem, and employ VQAs developed for optimization.
Along with the original Quantum Approximate Optimization Algorithm (QAOA) ansatz \cite{FarhiGoldstoneGutmann2014} and a variational form of an unstructured search, we will also examine the Quantum Multivariable Optimization Algorithm (QMOA), which was recently proposed as a VQA for the optimization of continuous multivariable functions \cite{MatwiejewPyeWang2023}.

This paper is organized as follows: Section~\ref{sec:background} will review the necessary background on gravitational wave detection and variational quantum algorithms.
In Section~\ref{sec:qalg}, we will provide details on the quantum computation of the objective function for the GW template search, which (along with the solution space encoding) is the main problem-dependent part of the VQAs.
We then describe the numerical simulations that were undertaken and present the results in Section~\ref{sec:numerics}.
This is followed by discussion in Section~\ref{sec:discussion} and the conclusion in Section~\ref{sec:conclusion}.

\section{Background}
\label{sec:background}

Here we will review some relevant background material on both gravitational wave detection and variational quantum algorithms.

\subsection{Gravitational wave detection}
\label{subsec:background_gwd}

This section will give a simplified account of the process of detecting gravitational wave signals in the output of the LIGO interferometers (see \cite{AllenEtal2012,UsmanEtal2016} for details on some of the actual search pipelines being used).
First, we will provide an overview of matched filtering, which is a general process used to look for a particular signal in noisy data.
Then we will discuss the GW signals that will be sought in the LIGO interferometer data.
Finally, we will review certain relevant aspects of the process of searching through a family of possible GW signals.

\subsubsection{Matched filtering}

Matched filtering is a data processing technique used to determine if a signal is present in noisy data.
It is applicable in cases where the form of the signal is known, as in the case of GW signals produced by compact binary coalescences (CBCs).

\begin{figure}[h]
  \centering
  \includegraphics[width=\linewidth]{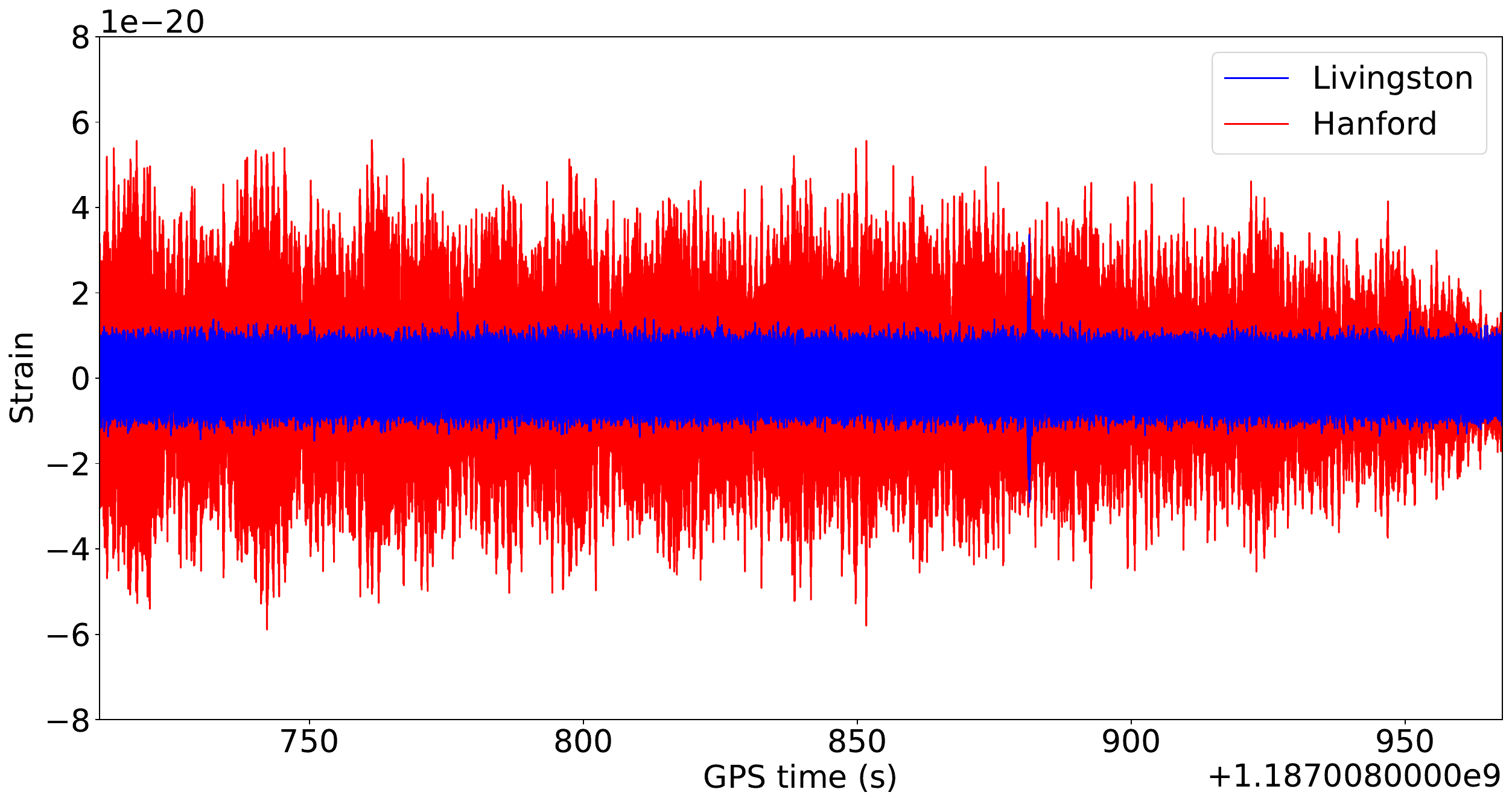}
  \caption{\small Strain data for the Livingston and Hanford detectors in a $256 \, \text{s}$ time interval (sampled at $4096 \, \text{Hz}$) containing the gravitational wave event GW170817 \cite{AbbottEtal2021}.}
  \label{fig:GW170817_timeseries}
\end{figure}

The output of the LIGO interferometers will be represented here as a time series of $N$ data points, indexed by $n \in \{ 0, \dots, N-1 \}$, which corresponds to a continuous signal sampled over some time interval, with sampling frequency $f_s$ (for example, see Figure~\ref{fig:GW170817_timeseries}).
The linear space of all possible time series over this interval is $\mathbb{R}^N$.
Much of the following analysis requires the use of the discrete Fourier transform of a time series $y = (y_n)_{n=0}^{N-1} \in \mathbb{R}^N$, defined as
\begin{equation}
  \tilde{y}_k := \Delta t \sum_{n=0}^{N-1} y_n e^{-2 \pi i \frac{nk}{N}},
\end{equation}
where $\Delta t := 1/f_s$.

If a signal is present in the output data, we will assume the time series $y = (y_n)_{n=0}^{N-1} \in \mathbb{R}^N$ can be decomposed as
\begin{equation}
  y_n = s_n + \nu_n,
\end{equation}
where $s_n$ is the time series of the idealized signal, and $\nu_n$ represents additive noise.
It is assumed that $\nu_n$ is a Gaussian random variable, with mean zero and covariance characterized by
\begin{equation}
  E[ \tilde{\nu}_k \tilde{\nu}_{k'}^\ast ] = \tfrac{1}{2\Delta f} S_k \delta_{kk'},
\end{equation}
where $S_k > 0$ and $E[\cdot]$ denotes the expectation value over noise realizations.
The sequence $(S_k)_{k=0}^{N-1}$ is called the noise \emph{power spectral density (PSD)}.
The factor of $1/(2 \Delta f)$ is included for consistency with the GW literature \cite{AllenEtal2012}, where $\Delta f := f_s / N$.

Next, we define a linear filter, $\xi$, as a linear map from time series to $\mathbb{R}$, which can generally be written as
\begin{equation}
  \xi(y) = \sum_{n=0}^{N-1} \xi_n y_n, \qquad \text{with } \xi_n \in \mathbb{R}.
\end{equation}
After applying some linear filter $\xi$ to the data $y$, the \emph{signal-to-noise ratio (SNR)} of the output can be defined by
\begin{equation}
\label{eq:snr_defn}
  \rho := \frac{\xi(y)}{\sqrt{E[|\xi(\nu)|^2]}}.
\end{equation}
The aim is then to design the linear filter $\xi$ so that the quantity $\rho$ can be used to indicate whether a particular signal $s$ is present in a data time series $y$.
Note that if the data contains only noise, $y_n = \nu_n$, then $E[\rho] = 0$, regardless of $\xi$.
If the data contains the signal, $y_n = s_n + \nu_n$, then $E[\rho] = \xi(s) / \sqrt{E[|\xi(\nu)|^2]}$.
To maximize our ability to distinguish between these two cases, the filter $\xi$ should be chosen to maximize the quantity $\xi(s) / \sqrt{E[|\xi(\nu)|^2]}$.
One can show this is achieved by
\begin{equation}
  \tilde{\xi}_k = \alpha \frac{\tilde{s}_k}{S_k},
\end{equation}
which is called the \emph{optimal} or \emph{matched filter}.
The constant $\alpha > 0$ can be chosen arbitrarily, since it does not affect the value of $\rho$.

The form of the matched filter suggests defining an inner product of two time series $x$ and $y$ as\footnote{Note that here we will use round brackets $(x|y)$ instead of angled brackets $\langle x | y \rangle$ to avoid confusing these with elements of the quantum state space in the following sections.}
\begin{equation}
\label{eq:ip}
  (x|y) := 2 \Delta f \sum_{k=0}^{N-1} \frac{\tilde{x}_k^\ast \tilde{y}_k}{S_k}.
\end{equation}
Note that this inner product is unitless.
For real-valued time series ($x_n^\ast = x_n$ and $y_n^\ast = y_n$), this quantity is real and can be simplified using $\tilde{x}_{N-k} = \tilde{x}_k^\ast$ and $\tilde{y}_{N-k} = \tilde{y}_k^\ast$.

Using this inner product, we can decompose the signal time series as
\begin{equation}
  s_n = A h_n,
\end{equation}
where $(h|h) = 1$, so that $A$ and $h$ describe the amplitude and ``shape'' of the signal, repectively.
The SNR of a time series $y$ after the matched filtering can then be expressed succinctly as
\begin{equation}
\label{eq:snr_ip}
  \rho = (h|y),
\end{equation}
since one can show that the denominator in \eqref{eq:snr_defn} becomes $E[|(h|\nu)|^2] = 1$.
The SNR is thus equal to the projection of the data onto the signal shape $h$, calculated using the inner product \eqref{eq:ip}, which is weighted by the inverse of the noise PSD.
The intuition behind this weighting is that a matching of the data to the signal in a region of frequency space where there are large noise fluctuations is a less reliable indicator of the presence of a signal than in regions with smaller noise fluctuations.
For data containing the signal, $y_n = A h_n + \nu_n$, the expectation value of the SNR is
\begin{equation}
  E[\rho] = E[(h| A h + \nu)] = A.
\end{equation}
Determining whether a particular signal, $s = A h$, is present in a time series, $y$, then amounts to computing the SNR \eqref{eq:snr_ip} and checking whether the result is greater than some predefined threshold, $\rho_0$.
This threshold is chosen to minimize the false alarm rate caused by noise fluctuations \cite{AllenEtal2012,UsmanEtal2016}.
On average, one should be able to use this procedure to detect signals with amplitudes $A > \rho_0$.

\subsubsection{GW templates and LIGO noise curves}

To use matched filtering for GW signal detection, we need knowledge of the noise PSD (for computing the inner product \eqref{eq:ip}) as well as the form of the time series corresponding to an idealized signal, called a \emph{template}.

\begin{figure}[h]
  \centering
  \includegraphics[width=0.6\linewidth]{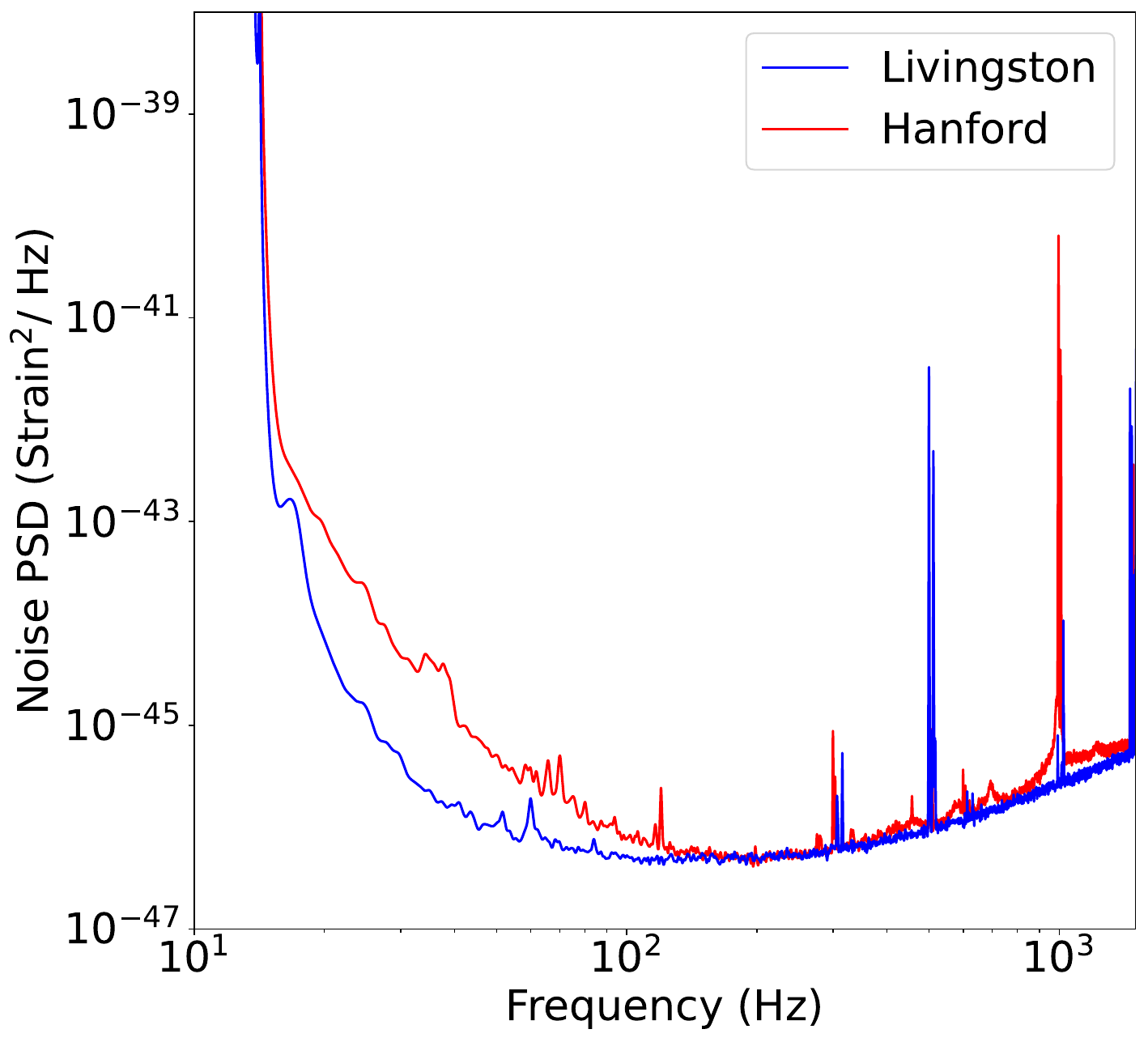}
  \caption{\small Noise PSDs obtained for the Livingston and Hanford detectors for a segment of strain data containing the event GW170817 \cite{AbbottEtal2021}. These curves were obtained using methods from the PyCBC library \cite{PyCBC}.}
  \label{fig:GW170817_psd}
\end{figure}

The noise PSD is estimated from the detector output using a variant of Welch's method (see \cite{AllenEtal2012} for details).
An example of the output of this procedure is shown in Figure~\ref{fig:GW170817_psd}.

In order to perform matched filtering, we also need to know the form of the signals.
For the coalescence of compact binaries (two black holes, a black hole and a neutron star, or two neutron stars), the dynamics of the binary and corresponding gravitational wave emission are predictions of the Einstein field equations of general relativity.
It is too difficult to solve these equations analytically for CBCs, however, it is possible to obtain approximations of the GW signals which are appropriate for the slow inspiral phase of two neutron stars or a neutron star and a black hole.
(Note that the binary black hole waveform can be obtained using numerical relativity, and the waveform is relatively short---typically a fraction of a second.)
These small mass systems with a slow inspiral phase are also those which dominate the number of templates required for the search problem, which motivates our focus on these systems in this work.

\begin{figure}[h]
  \centering
  \includegraphics[width=0.75\linewidth]{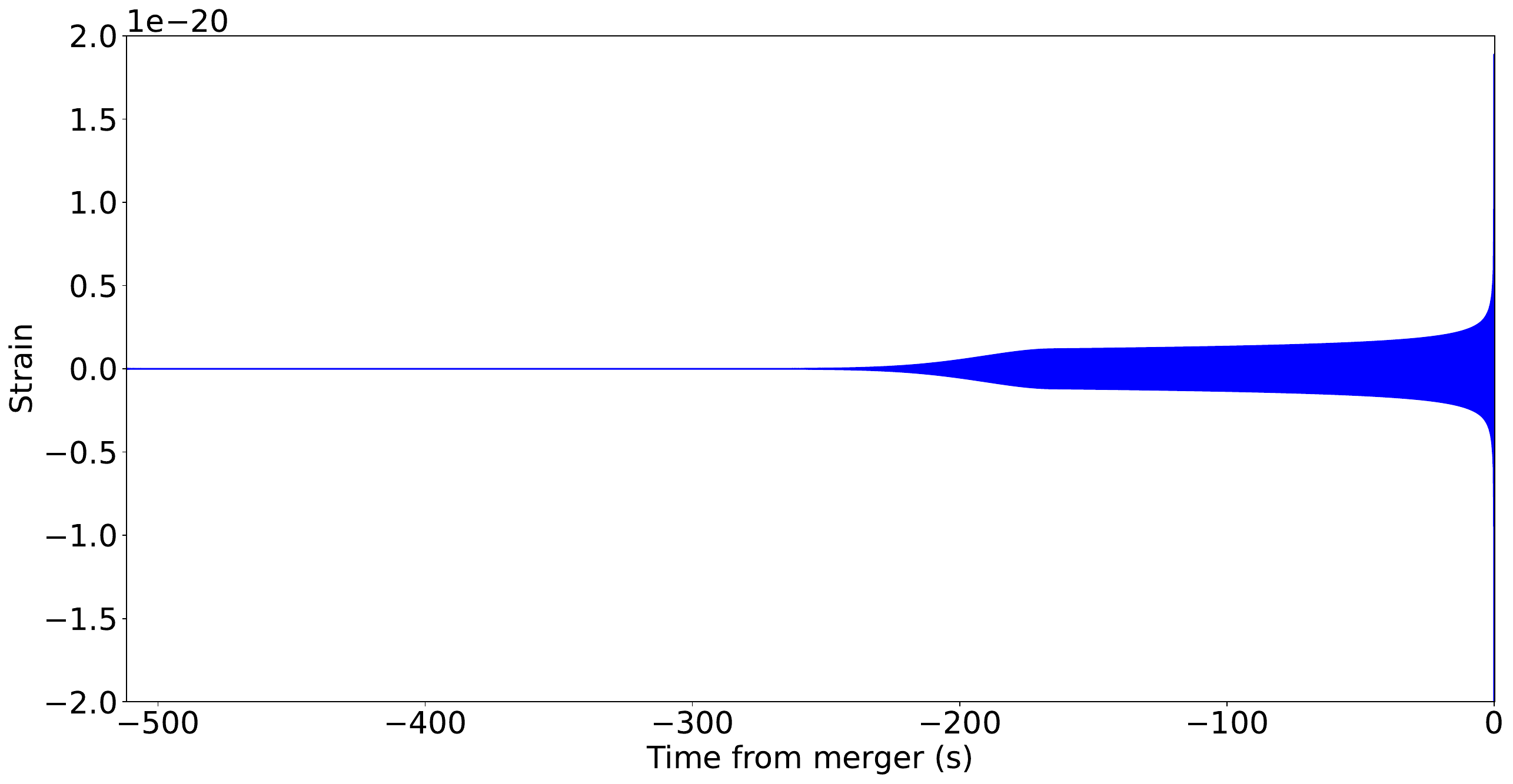}
  \caption{\small TaylorF2 template waveform in the time domain for the signal of the event GW170817. The mass parameters for this event are: $m_1 \approx m_2 \approx 1.3758 \, M_\odot$.}
  \label{fig:GW170817_template}
\end{figure}

If the total mass of the binary system is sufficiently small, $M \lesssim 12 \, M_\odot$, and effects of spin can be neglected, then a template for the GW signal can be written in frequency space as
\begin{equation}
\label{eq:template}
  \tilde{h}(f;m_1,m_2,t_c,\phi_c) = \mathcal{A} f^{-7/6} e^{i \left[ \frac{\pi}{4} + \phi_c - 2 \pi f t_c - \Psi(f;m_1,m_2) \right]} \qquad (f > 0),
\end{equation}
where $\mathcal{A}$ is a frequency-independent amplitude.
The template at negative frequencies is determined by $\tilde{h}(f) = \tilde{h}(-f)^\ast$.
The masses of the two components of the binary are denoted $m_1$ and $m_2$.
The parameters $t_c$ and $\phi_c$ are the time and phase of the waveform at coalescence.
The phase function (to 3.5-order in a restricted post-Newtonian (PN) expansion) is given by \cite{DamourIyerSathyaprakash2001,DamourIyerSathyaprakash2002,BuonannoEtal2009}
\begin{align}
\label{eq:template_phase}
  &\Psi(f;m_1,m_2) \nonumber \\
  &\quad = \frac{3}{128} \frac{1}{\eta v^5} \left[ 1 + \frac{20}{9} \left( \frac{743}{336} + \frac{11}{4} \eta \right) v^2 - 16 \pi v^3 + 10 \left( \frac{3058673}{1016064} + \frac{5429}{1008} \eta + \frac{617}{144} \eta^2 \right) v^4 \right. \nonumber \\
  &\quad + \left. \pi \left( \frac{38645}{756} - \frac{65}{9} \eta \right) \left( 1 + 3 \log(v/v_{lso}) \right) v^5 + \left( \frac{11583231236531}{4694215680} - \frac{640}{3} \pi^2 - \frac{6848}{21} \gamma \right. \right. \nonumber \\
  &\quad \left. - \left. \frac{6848}{21} \log(4v) + \left( - \frac{15737765635}{3048192} + \frac{2255 \pi^2}{12} \right) \eta + \frac{76055}{1728} \eta^2 - \frac{127825}{1296} \eta^3 \right) v^6 \right. \nonumber \\
  &\quad + \left. \pi \left( \frac{77096675}{254016} + \frac{378515}{1512} \eta - \frac{74045}{756} \eta^2 \right) v^7 \right]
\end{align}
where $\gamma \approx 0.5772\dots$ is the Euler-Mascheroni constant, $v := (\pi G M f / c^3)^{1/3}$ is a frequency-dependent dimensionless ``velocity'' parameter, $v_{lso} = 1/\sqrt{6}$, $M := m_1 + m_2$ is the total mass of the binary, and $\eta := m_1 m_2 / M^2$ is the symmetric mass ratio.
In the GW literature, this template is referred to as the TaylorF2 model \cite{DamourIyerSathyaprakash2001,DamourIyerSathyaprakash2002,BuonannoEtal2009} (see Figure~\ref{fig:GW170817_template}).
In this simplified model, there are only two \emph{intrinsic} parameters $(m_1,m_2)$ influencing the phase evolution of the waveform, and two remaining \emph{extrinsic} parameters $(t_c,\phi_c)$.
Of course, in principle there are many other quantities which affect the form of the signal: component spins, distance between source and detector, relative orientation, orbit eccentricity, etc..
In this model, these quantities are either assumed to be negligible (e.g., spins) or are absorbed into the overall amplitude $\mathcal{A}$.
Since the templates will be normalized for the matched filtering, we can ignore the form of $\mathcal{A}$.

The PN expansion of a GW emitted by a CBC can be thought of as the waveform associated with a binary orbit which is slowly decaying due to the emission of gravitational radiation.
The orbital velocity of the binary increases as the orbit decays, which increases the frequency of the emitted GWs, resulting in a ``chirp'' signal.
The PN expansion does not capture the dynamics of the merger or ringdown phases of the CBC, but \eqref{eq:template} provides a useful template for CBCs which have a long inspiral phase that dominate the GW signal (which is characteristic of low-mass systems).
Since the GW frequency increases as the inspiral phase progresses, the last stable orbit (lso) of the binary will determine a largest valid frequency for the template.
Typically this high-frequency cutoff is taken to be $f_{lso} = c^3 v_{lso}^3 / \pi G M$, corresponding to the velocity parameter of the innermost stable circular orbit on a Schwarzschild background, $v_{lso} = 1/\sqrt{6}$ \cite{BuonannoEtal2009,AllenEtal2012,CantonHarry2017}.
For binaries with $M \lesssim 4 \, M_\odot$, this corresponds to $f_{lso} \gtrsim 1 \, \text{kHz}$, which is at the edge of the sensitivity band of the detectors \cite{CantonHarry2017}.
Therefore, for small mass binaries, the power of the signal in the sensitivity band of the detectors primarily consists of the inspiral phase, and hence the template \eqref{eq:template} is useful for the detection of the GWs emitted by these systems.

Since the data analysis involves finite time and frequency series, we need to discretize the waveform \eqref{eq:template}.
The detector output time series corresponds to a continuous signal sampled at some rate $f_s$, with sample points $\{ t_n = t_0 + n/f_s \}_{n=0}^{N-1}$.
The frequency series are obtained by a discrete Fourier transform of these time series.
The frequency series associated with the above template is $\tilde{h}_k \approx f_s \tilde{h}(f_k) e^{2 \pi i t_0 f_k}$ (suppressing the parameter dependence) for $k \in \{ 1, \dots, \lfloor \tfrac{N-1}{2} \rfloor \}$, where $f_k := \tfrac{f_s}{N} k$.
The components $k \in \{ \lceil \tfrac{N-1}{2} \rceil + 1, \dots , N-1 \}$ are obtained from the real-valued time series constraint $\tilde{h}_k = \tilde{h}_{N-k}^\ast$.
For even $N$, we set $\tilde{h}_{N/2} \approx \text{Re} [ f_s \tilde{h}(f_{N/2}) e^{2 \pi i t_0 f_{N/2}} ]$.
The component $\tilde{h}_0$ can be omitted.
Note that the relation between $\tilde{h}_k$ and $\tilde{h}(f_k)$ is not exact, since sampling in time and then performing a discrete Fourier transform is not the same as sampling in frequency.
However, this is a good approximation for large $N$.
Therefore, the frequency series we will use for the templates is defined by
\begin{equation}
  \tilde{h}_k := \mathcal{N} \tilde{h}(f_k) \qquad \text{ for } k \in \{ 1, \dots, \lfloor \tfrac{N-1}{2} \rfloor \},
\end{equation}
and $\tilde{h}_{N/2} = \mathcal{N} \text{Re} [\tilde{h}(f_{N/2})]$ if $N$ is even.
The normalization constant is defined so that $(h|h) = 1$, and the phase $e^{2 \pi i t_0 f_k}$ from before can be omitted by absorbing $t_0$ into $t_c$ in \eqref{eq:template}.

Since the noise PSD rises sharply below $\approx 20 \, \text{Hz}$ (e.g., see Figure~\ref{fig:GW170817_psd}), often a low-frequency cutoff is introduced in the calculation of the inner product \eqref{eq:ip}.
We will use $f_L := 20 \, \text{Hz}$ as the cutoff, so that the frequency series will effectively begin at index $k_L := \max \{ 1, \lceil N f_L / f_s \rceil \}$.
Recall, we also have a high-frequency cutoff at $f_{lso}$, so we also write a maximum index as $k_H := \min \{ \lfloor \tfrac{N-1}{2} \rfloor, \lfloor N f_{lso} / f_s \rfloor \}$.
Therefore, the expression for the SNR associated with the template with parameters $(m_1,m_2,t_c,\phi_c)$ is
\begin{equation}
\label{eq:snr_with_extrinsic}
  \rho(m_1,m_2,t_c,\phi_c) = (h_{(m_1,m_2,t_c,\phi_c)}|y) = 4 \Delta f \hspace{1mm} \text{Re} \left[ \sum_{k=k_L}^{k_H} \frac{ \tilde{h}_k^\ast(m_1,m_2,t_c,\phi_c) \tilde{y}_k }{S_k} \right].
\end{equation}

The parameters $(m_1,m_2,t_c,\phi_c)$ corresponding to a particular GW signal in the data $y$ are not known a-priori.
Therefore, in order to establish whether a GW signal is present, we need to determine if there are any parameter values where the corresponding SNR exceeds the predefined threshold.

\subsubsection{Template search}

Searching through the set of possible template parameters is the step of the GW data analysis process which is the most computationally intensive.
This is the aspect which we would like to speed up using quantum algorithms.

Here we will formulate the problem of searching for parameters such that $\rho(m_1,m_2,t_c,\phi_c) > \rho_0$ as the problem of maximizing $\rho(m_1,m_2,t_c,\phi_c)$.
This is sufficient for solving the search problem, since we can then compare $\rho_{max}$ to $\rho_0$.
Any algorithm used to perform this optimization could include a stopping condition if a value of $\rho$ is found which exceeds $\rho_0$.

Fortunately, optimizing over the extrinsic parameters $(t_c, \phi_c)$ can be done efficiently.
In fact, $\phi_c$ can even be eliminated from the search.
Notice that, using \eqref{eq:snr_with_extrinsic} and \eqref{eq:template}, we can write
\begin{equation}
  \rho(m_1,m_2,t_c,\phi_c) = 4 \Delta f \hspace{1mm} \text{Re} \left[ e^{-i \phi_c} \sum_{k=k_L}^{k_H} \frac{ \tilde{h}_k^\ast(m_1,m_2,t_c,0) \tilde{y}_k }{S_k} \right].
\end{equation}
For each $(m_1,m_2,t_c)$, the sum in the above expression is some complex number, and $e^{-i \phi_c}$ simply changes the phase.
Then $\rho$ achieves its maximum when the phase is aligned with the positive real axis.
Since we are only seeking $\rho_{max}$, and not the parameter values which achieve this, then we can assume the $\phi_c$ optimization is already done by using the objective function
\begin{equation}
  \rho(m_1,m_2,t_c) = 4 \Delta f \left| \sum_{k=k_L}^{k_H} \frac{ \tilde{h}_k^\ast(m_1,m_2,t_c,0) \tilde{y}_k }{S_k} \right|.
\end{equation}
The parameter $t_c$ cannot be eliminated, but notice that
\begin{equation}
\label{eq:snr_simplified}
  \rho(m_1,m_2,t_c) = 4 \Delta f \left| \sum_{k=k_L}^{k_H} e^{2 \pi i f_k t_c} \frac{ \tilde{h}_k^\ast(m_1,m_2,0,0) \tilde{y}_k }{S_k} \right|.
\end{equation}
Therefore, for fixed $(m_1,m_2)$, instead of evaluating this sum for each (discrete) value of $t_c$ (requiring $\mathcal{O}(N^2)$ operations), we can use a fast Fourier transform (FFT) (requiring only $\mathcal{O}(N \log N)$ operations).

\begin{figure}[h]
  \centering
  \includegraphics[width=0.65\linewidth]{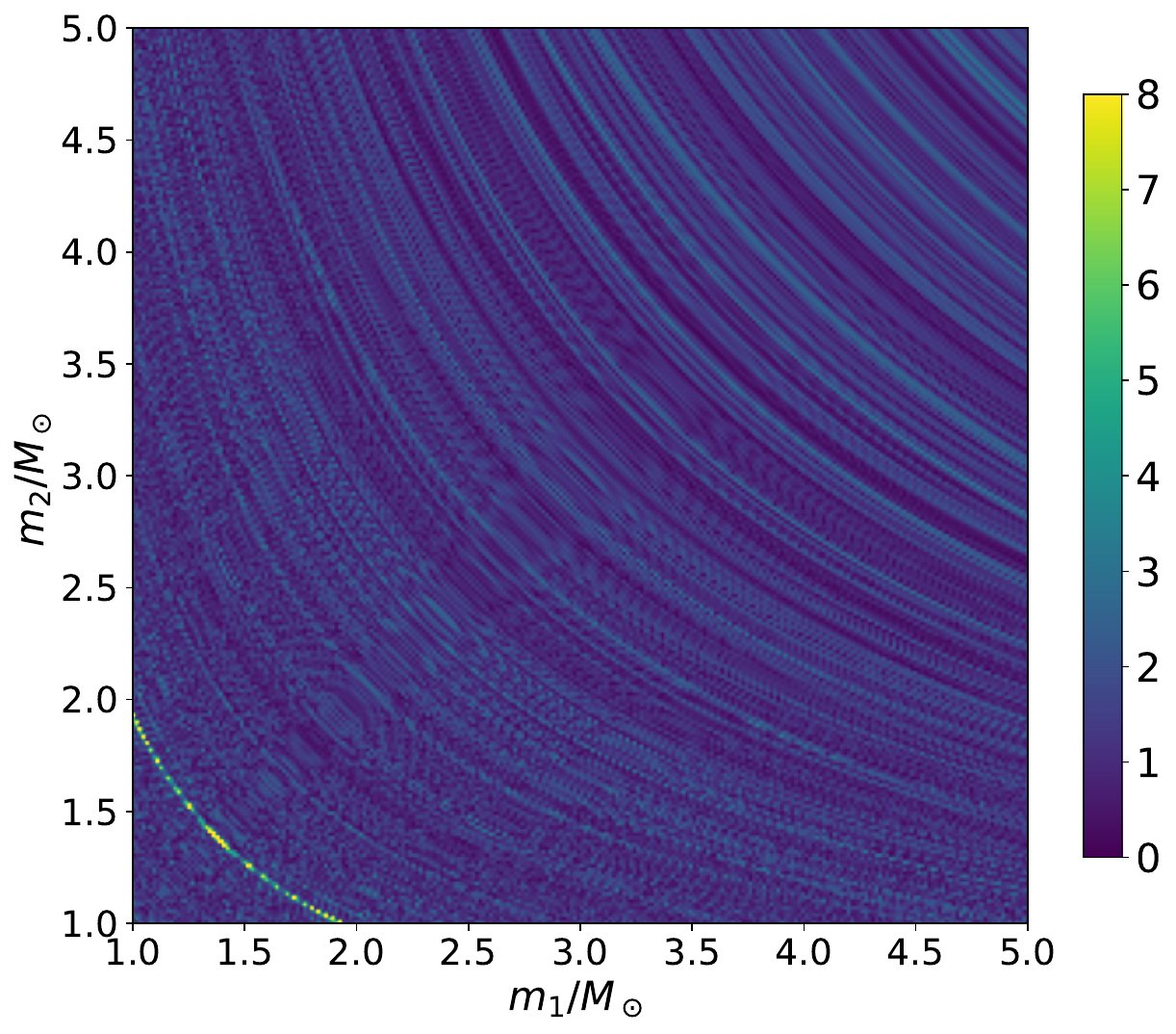}
  \caption{\small SNR $\rho$ as a function of the mass parameters $(m_1,m_2)$ for the strain data of the Livingston detector for the event GW170817 \cite{AbbottEtal2021}. The parameter $t_c$ was chosen to correspond to the time of coalescence. The $(m_1, m_2)$ parameters associated with the GW170817 signal, corresponding to the maximum of this function, are $m_1 \approx m_2 \approx 1.3758 \, M_\odot$. Although the maximum value of the SNR is $\rho \approx 26.2$, we set the colorbar to vary in the interval $[0,8]$ in order to better illustrate the features of the quality function below the later chosen threshold of $\rho_0 = 8$.}
  \label{fig:GW170817_snr_m1m2}
\end{figure}

Searching through the remaining intrinsic parameters $(m_1,m_2)$ is more difficult.
Typically this is done by discretizing the $(m_1,m_2)$ parameter space, and then finding the maximum of $\rho$ by grid search.
Figure~\ref{fig:GW170817_snr_m1m2} provides an illustration of the form of $\rho$ as a function of the two mass coordinates $m_1$ and $m_2$.
Determining an appropriate discretization of the mass parameter space is nontrivial, since one should ensure that there will always be a grid point sufficiently close to the parameters of any incoming GW signal to produce an SNR above the threshold \cite{Owen1996,OwenSathyaprakash1999}.
Typically the templates are placed in the parameter space so that the SNR of the closest template in the grid is at least 97\% of the SNR achieved with the true signal parameters.
This results in the search being dominated by the templates for low-mass systems, which is the regime where computational speedups would be most helpful (and the regime where the approximations \eqref{eq:template} and \eqref{eq:template_phase} are valid).
We will discuss our discretization scheme in Section~\ref{sec:qalg}.

\subsection{Variational quantum algorithms}
\label{subsec:background_vqa}

The term variational quantum algorithm (VQA) refers to an approach to quantum algorithm design in which a desired quantum operation is approximated by a predefined sequence of unitaries that depend on a set of classical parameters \cite{CerezoEtal2021}.
A classical optimization algorithm is then employed to take samples of the output of the quantum circuit and tune these parameters toward the desired operation.
The quantum algorithm (consisting of initialization, execution of unitaries, and measurement) is repeated after each step of the classical optimization algorithm to generate new samples with the updated parameters.
For NISQ devices, repeated iteration of a low-depth quantum circuit is preferable to attempting a deeper circuit where errors accumulate to the extent that the output is too noisy to be useful.
Despite the use of shallow quantum circuits, and the overhead of the classical optimization and repetitions of the quantum algorithm, VQAs show promise for being able to provide a quantum advantage for certain problems \cite{FarhiGoldstoneGutmann2014,PeruzzoEtal2014,Preskill2018,CerezoEtal2021}.

In this paper, we consider VQAs which aim to compute the maximum (to some approximation) of a given real-valued objective function, $f$, over a finite set $j \in \{ 0, \dots, J-1 \}$.
For simplicity, we will assume $J = 2^q$ for some $q \in \mathbb{N}$, so that each element in the domain of $f$ can be identified with a computational basis state, $\{ \ket{j} \}_{j=0}^{J-1}$, of a quantum system of $q$ qubits.
The VQAs we will consider here will begin with an equal superposition over these basis states,
\begin{equation}
  \ket{\psi_0} := \frac{1}{\sqrt{J}} \sum_{j=0}^{J-1} \ket{j}.
\end{equation}
The unitary ansatz will have the general structure of alternating phase shifts and mixing operations,
\begin{equation}
\label{eq:vqa_ansatz}
  \hat{U}(\bm{t},\bm{\gamma}) := \prod_{\ell=1}^p e^{-i t_\ell \hat{W}} e^{-i \gamma_\ell \hat{Q}},
\end{equation}
where $p$ is a fixed depth,
\begin{equation}
  \hat{Q} := \sum_{j=0}^{J-1} f(j) \ket{j} \bra{j}
\end{equation}
is a diagonal operator encoding the classical function we want to maximize, $\hat{W}$ is the generator of the mixing unitary (which we will discuss further below), and $\bm{t} = (t_\ell)_{\ell=1}^p$ and $\bm{\gamma} = (\gamma_\ell)_{\ell=1}^p$ are the classical variational parameters.
The final state,
\begin{equation}
  \ket{\bm{t},\bm{\gamma}} := \hat{U}(\bm{t},\bm{\gamma}) \ket{\psi_0},
\end{equation}
is then measured in the computational basis to provide a sample for the estimation of
\begin{equation}
  \langle \bm{t}, \bm{\gamma} | \hat{Q} | \bm{t}, \bm{\gamma} \rangle = \sum_{j=0}^{J-1} f(j) | \langle j | \bm{t}, \bm{\gamma} \rangle |^2.
\end{equation}
This quantity is then given to the classical optimizer, which aims to maximize it (and hence $f$) by varying $(\bm{t},\bm{\gamma})$.

The phase shift, $e^{-i \gamma \hat{Q}}$, is straightforward to construct for functions, $f$, which can be computed by an efficient quantum algorithm.
That is, we suppose that we have access to a unitary which maps $\ket{j}\ket{0} \to \ket{j}\ket{f(j)}$.
The second quantum register here consists of $q'$ qubits, with computational basis states representing a discretization of some interval $(a,b)$ of the codomain of $f$ to $q'$ bits of precision (with $(a,b)$ chosen large enough to avoid overflow errors).
Given such a unitary, the phase shift can be implemented by
\begin{align}
  \ket{j}\ket{0} &\to \ket{j}\ket{f(j)} \nonumber \\
  &\to \left( \hat{I} \otimes e^{-i \gamma \left[ a + \frac{(b-a)}{(2^{q'}-1)} \sum_{k=1}^{q'} 2^{k-2} ( \hat{I} - \hat{Z}^{(k)} )  \right]} \right) \ket{j}\ket{f(j)} = e^{-i \gamma f(j)} \ket{j}\ket{f(j)} \nonumber \\
  &\to e^{-i \gamma f(j)} \ket{j}\ket{0},
\end{align}
where the final step reverses the computation of $f$, which will disentangle the registers when the input is in a superposition (after this step the $q'$ working qubits can be ignored).
The operator $\hat{Z}^{(k)}$ denotes the Pauli-$Z$ operator acting on the $k^{th}$ qubit of the second register.
Thus, the phase shift can be easily implemented with the unitary computing $f$ and a product of single-qubit rotations.

The mixing operation, $e^{-i t \hat{W}}$, has the role of changing the probabilities of measuring the states $\{ \ket{j} \}_{j=0}^{J-1}$, with the aim of increasing the probability of measuring $j$ which maximizes $f$.
Following \cite{MarshWang2019,MarshWang2020}, we will consider mixing operations which correspond to continuous-time quantum walks, with
\begin{equation}
\label{eq:mixing_operator}
  \hat{W} := \sum_{j,j'=0}^{J-1} w_{jj'} \ket{j} \bra{j'},
\end{equation}
where $w_{jj'} \in \{0,1\}$ is identified with the adjacency matrix of some chosen graph with $J$ vertices (then $w_{jj'} = 1$ if and only if $j$ and $j'$ are connected by an edge).

The first proposed VQA of the general form \eqref{eq:vqa_ansatz} was the Quantum Approximate Optimization Algorithm (QAOA) used for the task of combinatorial optimization \cite{FarhiGoldstoneGutmann2014}.
This algorithm uses the mixing operator, $\hat{W} = \sum_{k=1}^q {\hat{X}^{(k)}}$, which is easily implemented as a product of single-qubit rotations, and can be shown corresponds to the adjacency matrix of a hypercube graph (where two bit strings are connected by an edge if and only if their Hamming distance is 1).
Extensions of QAOA have been proposed to handle constrained optimization \cite{HadfieldEtal2019,MarshWang2019,MarshWang2020}, but we will not need to consider constraints in this paper.

Another notable quantum algorithm which is comparable to these VQAs is Grover's quantum search algorithm \cite{Grover1996}.
To make this comparison, one can think of a search problem as a maximization of a binary objective function, which takes the value $f(j) = 1$ if $j$ is a solution to the search problem, and $f(j) = 0$ otherwise.
The unitary operator used in Grover's algorithm is of the form \eqref{eq:vqa_ansatz}, with $\hat{W}$ corresponding to the adjacency matrix of a complete graph, $t_\ell = \pi/J$ and $\gamma_\ell = \pi$ are fixed $\forall \ell$, and $p = \lceil \frac{\pi}{4 \text{asin}\sqrt{J_S/J}} - \frac12 \rfloor \approx \frac{\pi}{4} \sqrt{J/J_S}$ (where $J_S$ is the number of elements $j$ such that $f(j) = 1$, and where $J \gg J_S$).
As opposed to VQAs, the parameters in Grover's algorithm are predetermined\footnote{If $J_S$ is not known a-priori, it can be determined with a $\mathcal{O}(\sqrt{J})$ \emph{quantum counting} algorithm.}, and it is guaranteed to converge to a solution of the search problem (with success probability $\geq 1 - \frac{J_S}{J}$) at a predetermined depth $p \sim \sqrt{J/J_S}$ (which is a quadratic speedup over a classical search).
However, for large search spaces ($J \gg J_S$), a depth of $p \sim \sqrt{J/J_S}$ is beyond the capabilities of NISQ devices.
For quantum computers which can only perform circuits of a limited depth, we can employ a restricted-depth Grover's search (RDGS), which simply executes the unitary operations employed in Grover's algorithm up to a fixed depth $p$.
The probability of measuring some solution $j_S$ to the search problem at depth $p$ is
\begin{equation}
  \text{Pr}(j \in \{ j_S \} ) = \sin^2 \left[ (2p+1) \text{asin}\sqrt{J_S/J} \right].
\end{equation}

It is well-known that the full Grover search algorithm is optimal for an unstructured quantum search (i.e., if we have no prior knowledge about the binary objective function and can only learn about it through the application of the phase shift operation) \cite{Zalka1999}.
Therefore, the speedup provided by a restricted-depth Grover search gives an upper bound on the performance of VQAs for unstructured search problems.
However, one can then ask if VQAs can be more efficient than Grover's algorithm for certain \emph{structured} optimization problems.
Notice that the use of a complete graph for the mixing operation in Grover's algorithm removes any possible structure present in an objective function, since the complete graph connects every element in the search space to each other.
This suggests that one should look at different ways of connecting the search space (i.e., choosing different graphs for $\hat{W}$), in order to reflect some particular structure in the problem \cite{AaronsonAmbainis2009,McCleanEtal2021}.

Recently, a graph structure was proposed for the optimization of continuous multivariable functions within the VQA framework, called the Quantum Multivariable Optimization Algorithm (QMOA) \cite{MatwiejewPyeWang2023}.
Consider an objective function of multiple variables, $f(j_1, \dots, j_D)$, with $j_d \in \{ 0, \dots, J_d-1 \}$ where $d \in \{ 1, \dots, D \}$.
Of course, this search space can also be indexed by a single variable $j \in \{ 0, \dots, \prod_{d=1}^D J_d - 1 \}$, but the idea of QMOA is that maintaining this separation can be useful to reflect the structure of certain optimization problems.
For example, the objective function may correspond to the discretization of a continuous multivariable function, where each index $j_d$ corresponds to a value of one of the discretized coordinates of the domain of the function.
The elements of the search space are represented in a system of $\log_2 \prod_{d=1}^D J_d$ qubits by computational basis states $\{ \ket{ j_1, \dots, j_D } \}_{j_d=0}^{J_d-1}$.
The mixing operator of QMOA is of the form
\begin{equation}
\label{eq:qmoa_mixer}
  e^{-i \sum_{d=1}^D t_d \hat{W}_d} = e^{-i t_1 \hat{W}_1} \otimes \cdots \otimes e^{-i t_D \hat{W}_D},
\end{equation}
where each $\hat{W}_d$ acts on the subsystem corresponding to the variable $j_d$, and is taken to correspond to the adjacency matrix of a circulant graph.
The use of circulant graphs allows for an efficient implementation of the mixing operation by use of the quantum Fourier transform \cite{QiangEtal2016,LokeWang2017,ZhouWang2017}.
Note also that this algorithm allows for the independent parametrization of the walk times for each variable, which means there are $(D+1)p$ variational parameters at depth $p$, rather than only $2p$.
In \cite{MatwiejewPyeWang2023}, it was demonstrated numerically that QMOA is more effective than a restricted-depth Grover's search for the optimization of a diverse set of continuous multivariable functions.
Here we want to examine how effectively QMOA, as well as other VQAs, are able to perform for the GW template search, as compared to the restricted-depth Grover search.

\section{Quantum algorithm for GW template search}
\label{sec:qalg}

The key aspect to discuss in using VQAs for the GW template search problem is the computation of the quality function $\ket{j}\ket{0} \to \ket{j} \ket{f(j)}$ on a quantum computer.
The remaining steps of a VQA then proceed as described above, as they are largely problem-independent.

For the GW search problem, we want to maximize the function 
$\rho(m_1,m_2,t_c)$ in \eqref{eq:snr_simplified} for a given time series, $y$, and noise PSD, $S$, both of which are classical inputs.
The SNR, $\rho(m_1,m_2,t_c)$ is then a function of three parameters.
In this work, we will focus only on the search over the mass parameters, since this is the aspect of the problem we are interested in speeding up.
For the numerical simulations in Section~\ref{sec:numerics}, we will use a time series $y$ where the time of coalescence $t_c$ is known, and simply use this fixed value in our computation of $\rho(m_1,m_2,t_c)$.

The next step is to discretize the parameter space $(m_1,m_2)$.
As we mentioned at the end of Section~\ref{subsec:background_gwd}, this should be done to make sure that the SNR of the closest grid point to the parameters of any incoming signal has at least 97\% of the SNR when matched with the exact parameters of the signal.
There are a number of algorithms which can be used for such template placement \cite{CantonHarry2017,CapanoEtal2016}, however, here we will not attempt to develop quantum versions of these algorithms.
For this preliminary work, we will instead construct a simpler discretization of the parameter space which approximates the actual grids used.
For instance, our grid will retain the feature that there is a higher concentration of templates for low-mass systems.

\begin{figure}[h]
  \centering
  \includegraphics[width=0.65\linewidth]{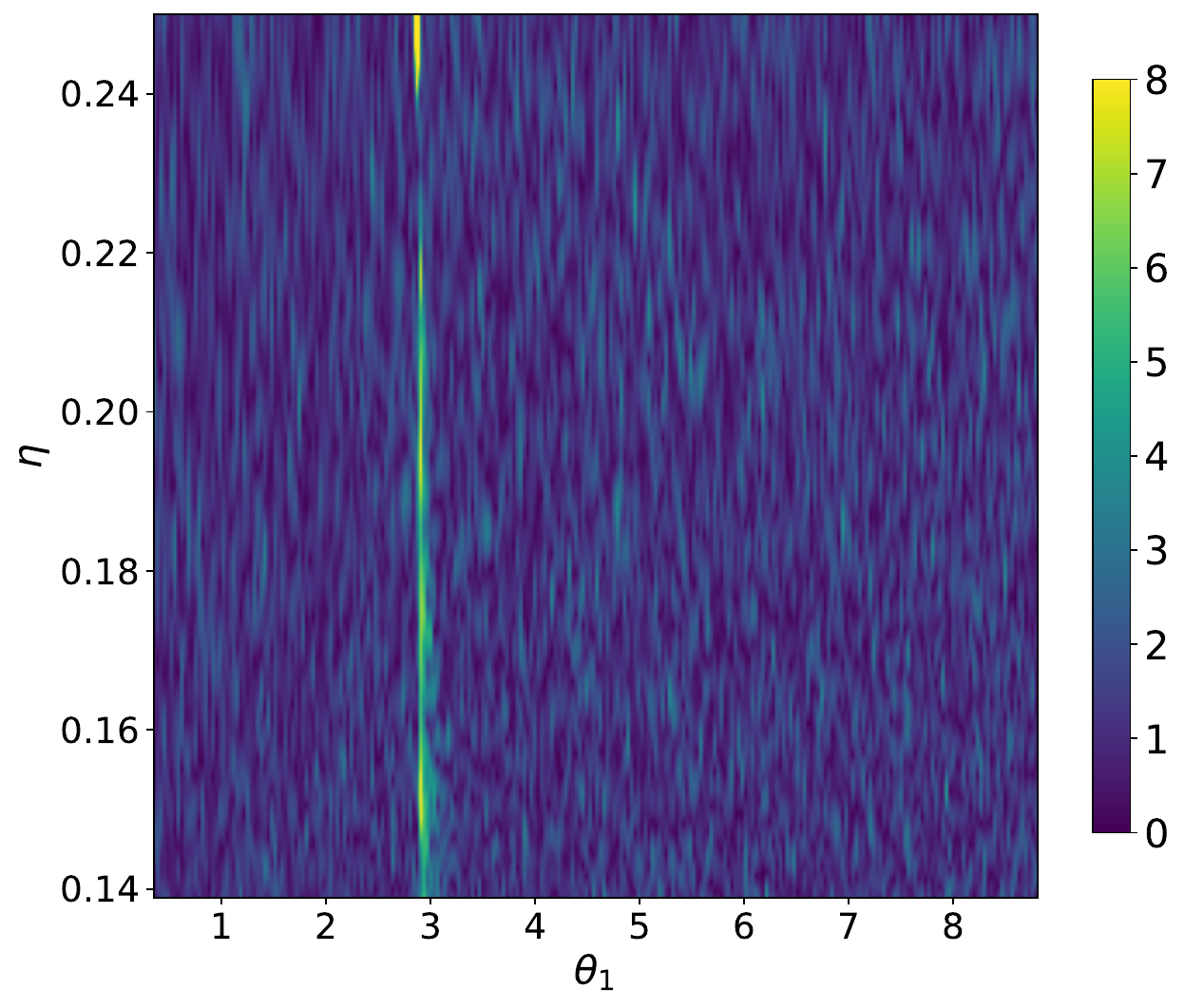}
  \caption{\small SNR $\rho$ as a function of the mass parameters $(\theta_1,\eta)$ for the strain data of the Livingston detector for the event GW170817 \cite{AbbottEtal2021}. The parameter $t_c$ was chosen to correspond to the time of coalescence. The $(\theta_1, \eta)$ parameters associated with the GW170817 signal, corresponding to the maximum of this function, are $\theta_1 \approx 2.87$ and $\eta \approx 0.25$. As in Figure~\ref{fig:GW170817_snr_m1m2}, the colorbar is scaled to illustrate the features of the quality function below the threshold of $\rho_0 = 8$.}
  \label{fig:GW170817_snr_theta1eta}
\end{figure}

To do this, we first change the mass coordinates to the symmetric mass ratio $\eta = m_1 m_2 / M^2$ (where $M = m_1 + m_2$) and the (dimensionless) Newtonian chirp time \cite{OwenSathyaprakash1999},
\begin{equation}
  \theta_1 := \frac{5}{128} (\pi G M f_s / c^3)^{-5/3} \eta^{-1},
\end{equation}
which is closely related to the \emph{chirp mass}, $\mathcal{M} := M \eta^{3/5} \sim \theta_1^{-3/5}$.
Figure~\ref{fig:GW170817_snr_theta1eta} shows the SNR $\rho(\theta_1,\eta)$ expressed in these coordinates.
To create the grid, we first choose intervals $\theta_1 \in (\theta_{1,\text{min}}, \theta_{1,\text{max}})$ and $\eta \in (\eta_{\text{min}}, \eta_{\text{max}})$ to cover the parameter region $1 \, M_\odot \leq m_1, m_2 \leq 5 \, M_\odot$, and then uniformly discretize these intervals.
With this uniform grid in the $(\theta_1,\eta)$ coordinates, there is a higher concentration of templates for low mass systems, which can be seen from $dM = -\tfrac35 C^{-1} M^{8/3} \eta d\theta_1 - \tfrac35 M \eta^{-1} d\eta$ with $C \equiv \tfrac{5}{128} (\pi G f_s / c^3)^{-5/3}$.
As mentioned earlier, it could happen that the mass parameters of an incoming signal lie between grid points in the discretized parameter space.
For the numerical simulations in the next section, we will address this by simply shifting the $\theta_1$ and $\eta$ intervals so that one of the grid points corresponds exactly to the parameters of the signal known to exist in the corresponding time series.

For quantum optimization over $(\theta_1,\eta)$, we use qudit registers to store discretized intervals of values for each parameter.
Explicitly, a register of $q_1$ qubits used to represent $\theta_1$ on the interval $(\theta_{1,\text{min}}, \theta_{1,\text{max}})$ is written as
\begin{equation}
  \ket{\theta_1} \quad \text{ with } \theta_1 \in \bigl\{ \tfrac{(\theta_{1,\text{max}} - \theta_{1,\text{min}})}{(2^{q_1}-1)} j_1 + \theta_{1,\text{min}} \bigr\}_{j_1 = 0}^{2^{q_1}-1}.
\end{equation}
We similarly use a register of $q_2$ qubits to represent $\ket{\eta}$ for a uniform discretization of $\eta$ on an interval $(\eta_{\text{min}}, \eta_{\text{max}})$.
We then have a basis of quantum states, with each element of the basis representing one of the $J = 2^{q_1} \times 2^{q_2}$ grid points in the parameter space.

\begin{figure}[h]
  \centering
  \includegraphics[width=0.65\linewidth]{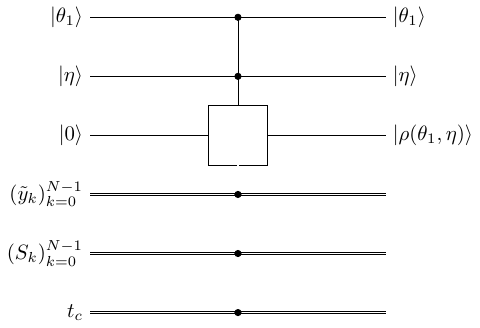}
  \caption{\small Illustration of the quantum operation $\ket{\theta_1,\eta} \ket{0} \to \ket{\theta_1,\eta} \ket{\rho(\theta_1,\eta)}$. The quantities $\tilde{y}$, $S$, and $t_c$ are classical inputs to the computation (indicated by double lines).}
  \label{fig:qcircuit_quality}
\end{figure}

Computing $\ket{\theta_1,\eta} \ket{0} \to \ket{\theta_1,\eta} \ket{\rho(\theta_1,\eta)}$ then requires computing the template for the $(\theta_1,\eta)$ parameters, and combining this with the classical inputs $\tilde{y}$, $S$, and $t_c$ to compute the sum in \eqref{eq:snr_simplified}.
In a classical GW search, the template frequency series $( \tilde{h}_k(\theta_1,\eta))_{k=k_L}^{k_H}$ is often precomputed and stored in a \emph{template bank}.
Storing a template bank in a quantum computer could be done in principle using QRAM \cite{GiovannettiLloydMaccone2008}.
However, in \cite{GaoEtal2022} it was suggested that instead of using QRAM, one could compute the template directly from the mass parameter values on the quantum computer.
For many GW signals, such as those examined in \cite{GaoEtal2022}, this would involve doing numerical relativity on a quantum computer, which is not likely to be feasible until we have very large fault-tolerant quantum computers.
For GWs produced by small mass CBCs, however, computing $\tilde{h}_k(\theta_1,\eta)$ from $(\theta_1,\eta)$ (for each $k$) only requires the computation of elementary mathematical functions given in \eqref{eq:template} and \eqref{eq:template_phase}.
Since it can be computed efficiently classically, it follows that can be done efficiently on a quantum computer as well.
Due to the relatively simply structure of the function $\rho(\theta_1,\eta)$ for small mass CBCs, the GW template search in this regime is perhaps suitable as a near-term application of quantum computers.
Figure~\ref{fig:qcircuit_quality} provides a quantum circuit illustration of the general structure of the operation $\ket{\theta_1,\eta} \ket{0} \to \ket{\theta_1,\eta} \ket{\rho(\theta_1,\eta)}$.

Note that computing $\tilde{h}_k(\theta_1,\eta)$ for each $k$ and evaluating the sum over $k$ in the expression in \eqref{eq:snr_simplified} for $\rho$ requires $\mathcal{O}(N)$ operations, where recall $N$ is the number of elements in the time series $y$.
For the template search problem, we are primarily interested in efficiency in the number of templates we are searching through, which we denoted $J$ (i.e., the number of points in the parameter grid).
Therefore, the important point is that computing $\ket{\theta_1,\eta} \ket{0} \to \ket{\theta_1,\eta} \ket{\rho(\theta_1,\eta)}$ can be done efficiently in $q_1$ and $q_2$, with $q_1 + q_2$ being the number of qubits in the registers $\ket{\theta_1,\eta}$.

\section{Numerical simulations}
\label{sec:numerics}

Here we describe and present the results of classical numerical simulations undertaken to assess the performance of various VQAs for the optimization problem arising in the GW matched filtering problem.
All of the code necessary to reproduce these results is openly available at \cite{Matwiejew2022}.

\subsection{Methods}

We obtained GW open science data from LIGO through \cite{AbbottEtal2021} and used signal processing methods from the PyCBC library \cite{PyCBC}.
Our simulations were mainly focused on a particular $256 \, \text{s}$ time segment containing the signal event GW170817.
The GW signal in this event has a relatively large SNR ($\rho \approx 26.2$), and corresponds to a binary neutron star merger with both neutron star masses equal to approximately $m_1 \approx m_2 \approx 1.3758 \, M_\odot$.
For simplicity, we only used data from the Livingston detector (in which there is a detector glitch near the time of the merger, but a cleaned version of the data, which was used here, is available through \cite{AbbottEtal2021}).
The noise PSD was obtained from this segment of data using methods from the PyCBC library \cite{PyCBC}.
Using the above mass parameters for this signal, we found that for this $256 \, \text{s}$ segment of data, $t_c \approx 170.7 \, \text{s}$ is the value which maximizes \eqref{eq:snr_simplified}.
We thus fix this value of $t_c$ throughout the simulations.

Unless otherwise stated, the $(\theta_1,\eta)$ parameter space was discretized using a uniform $256 \times 256$ grid (corresponding to $\log_2 256^2 = 16$ qubits) covering the region $1 \, M_\odot \leq m_1, m_2 \leq 5 \, M_\odot$.
The grid was then shifted so that the GW170817 signal parameters corresponded exactly to one of the grid points.

The classical simulations of the VQAs were performed using the QuOp\_MPI library \cite{MatwiejewWang2022,Matwiejew2022}.
For the optimization of the variational parameters, we used the default setting of QuOp\_MPI, which uses the SciPy v1.11.4 implementation of the Broyden-Fletcher-Goldfarb-Shanno (BFGS) algorithm via its \texttt{minimize} function (with a gradient tolerance set to $10^{-3}$) \cite{MatwiejewWang2022,Matwiejew2022,SciPy}.
Due to the stochastic nature of the optimization, we repeated the simulation of each of the VQAs 10 times.
The results presented below show the average and standard deviation over these 10 repeats.

In our primary simulation, we examined the performance of four VQAs (i.e., VQAs with different mixing unitaries).
The first is the original QAOA, with the mixing operator \eqref{eq:mixing_operator} corresponding to the adjacency matrix of a hypercube graph, which we will denote as QAOA (Hypercube).
For the second, we take \eqref{eq:mixing_operator} to be the adjacency matrix of a complete graph, which we call QAOA (Complete).
We also examine the performance of two variants of QMOA for the two-dimensional optimization over $(\theta_1,\eta)$.
That is, in \eqref{eq:qmoa_mixer}, $D=2$, $e^{-i t_1 \hat{W}_1}$ acts only on the register $\ket{\theta_1}$, and $e^{-i t_2 \hat{W}_2}$ acts on $\ket{\eta}$.
The variants of QMOA we simulated are for $\hat{W}_1$ and $\hat{W}_2$ corresponding to adjacency matrices of complete graphs, called QMOA (Complete), and of cycle graphs, called QMOA (Cycle).
We compare the performance of these VQAs to a restricted-depth Grover search (RDGS) (as described in Section~\ref{subsec:background_vqa}), where the binary objective function is taken to be $1$ if $\rho(\theta_1,\eta)$ is above a threshold of $\rho_0 = 8$, and $0$ otherwise.
An SNR threshold of $\rho_0 = 8$ is a typical value chosen in GW signal detection \cite{CapanoEtal2016}.
For the event GW170817, the SNR is $\rho \approx 26.2$ for the Livingston detector.

To assess the performance of the VQAs, we are primarily interested in the probability that a computational-basis measurement of the final wavefunction $\ket{\bm{t},\bm{\gamma}}$ returns parameter values $(\theta_1,\eta)$ with an SNR, $\rho(\theta_1,\eta)$, above the threshold.
That is,
\begin{equation}
  \text{Prob}[\hat{Q} > \rho_0] = \sum_{\{ \theta_1,\eta \text{ $:$ } \rho(\theta_1,\eta) > \rho_0 \}} | \langle \theta_1,\eta | \bm{t}, \bm{\gamma} \rangle |^2,
\end{equation}
where the sum is over the points in the discretized parameter grid with $\rho(\theta_1,\eta) > \rho_0$.
This corresponds to the probability that the signal present in the time series is successfully identified.
In addition, we will also report the expectation value of the quality function,
\begin{equation}
  \langle \hat{Q} \rangle = \sum_{\theta_1,\eta}  \rho(\theta_1,\eta) | \langle \theta_1,\eta | \bm{t}, \bm{\gamma} \rangle |^2.
\end{equation}
We are interested in how these quantities behave as functions of the depth $p$ of the VQAs, which corresponds to the number of times the function $\rho(\theta_1,\eta)$ is queried in the quantum algorithms. 

\subsection{Results}

\begin{figure}[ht]
  \centering
  \begin{subfigure}{0.45\linewidth}
    \centering
    \includegraphics[width=\linewidth]{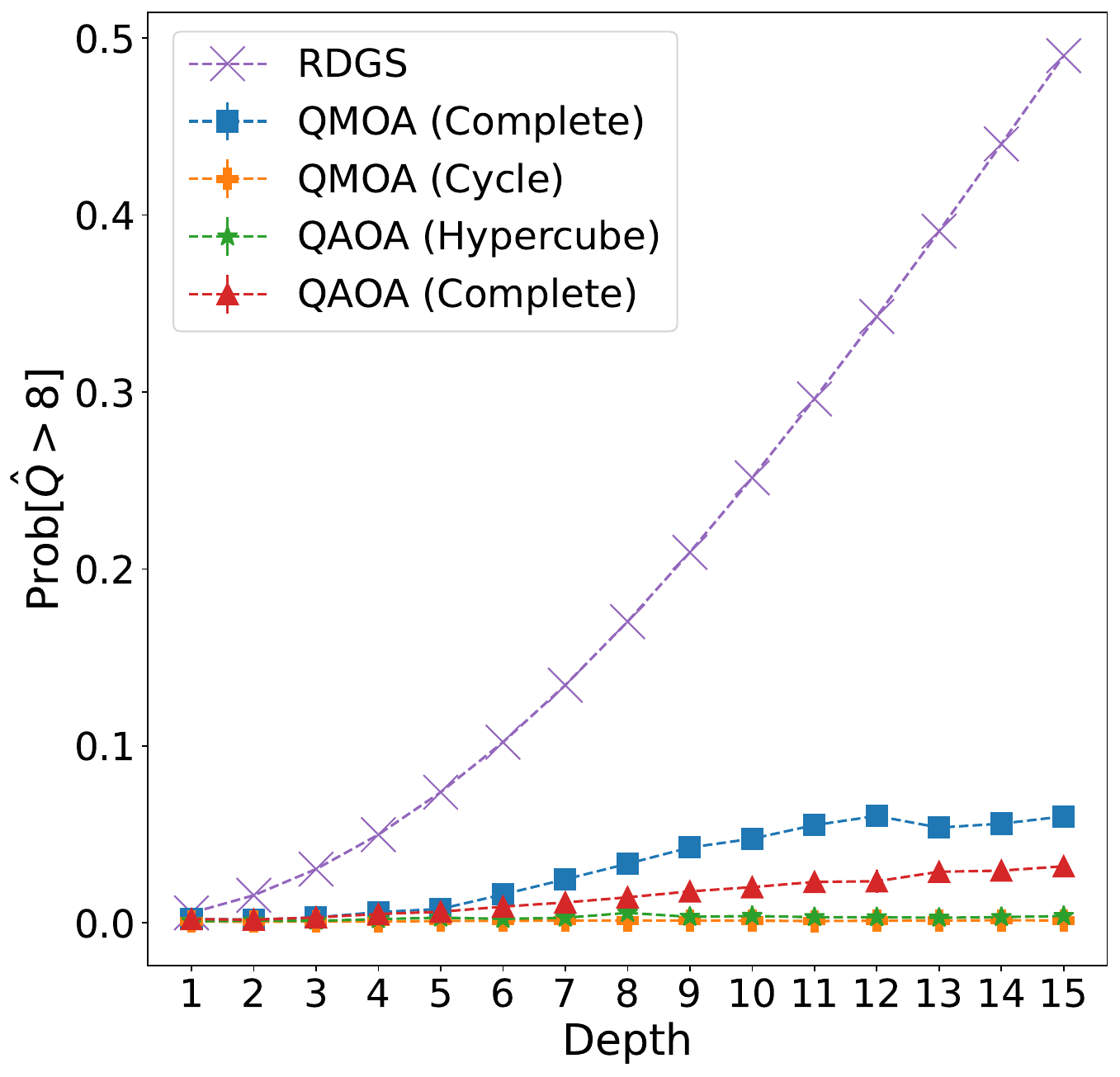}
    \caption{\small Probability of success}
  \end{subfigure}
  \begin{subfigure}{0.45\linewidth}
    \centering
    \includegraphics[width=0.95\linewidth]{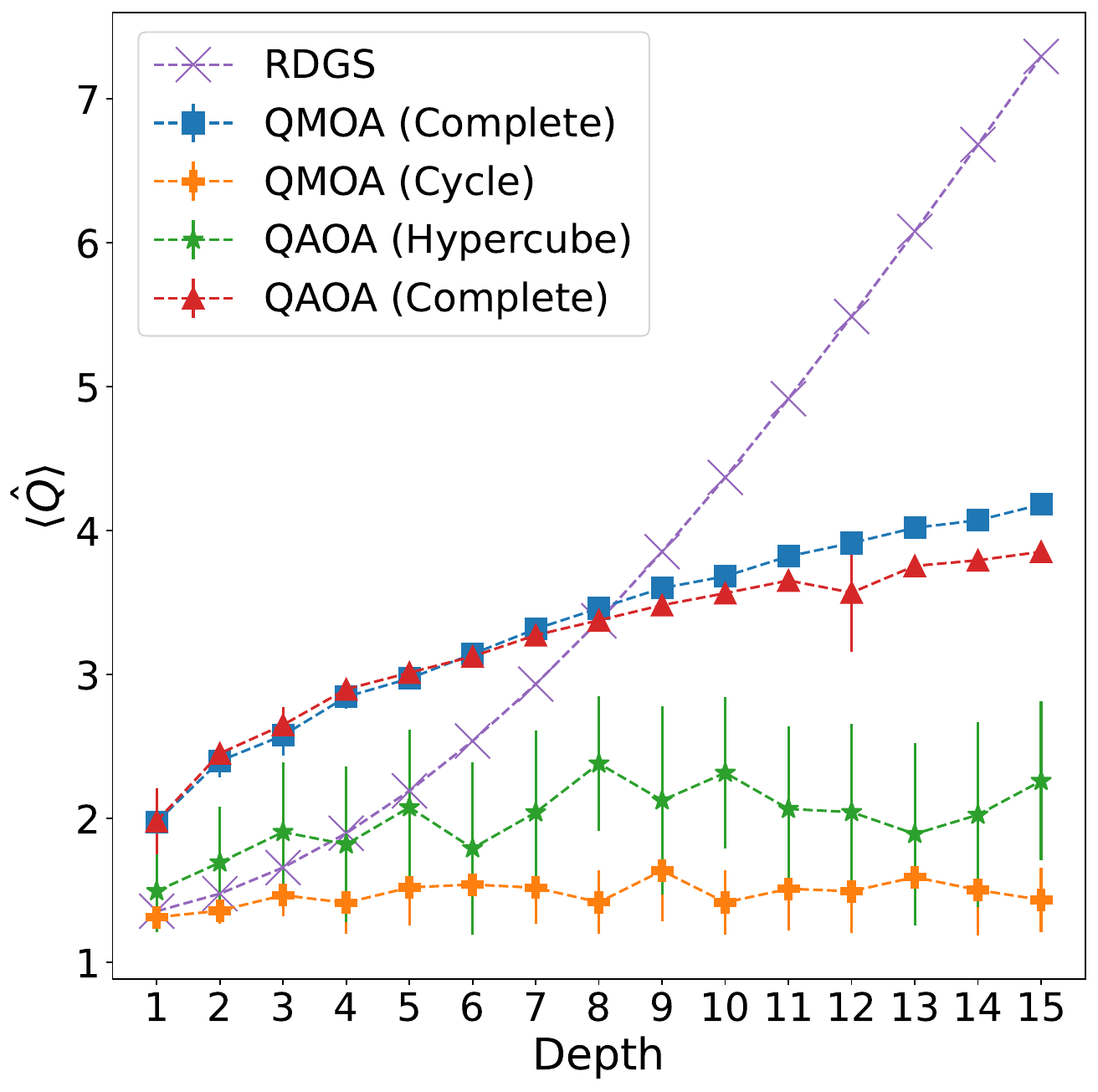}
    \caption{\small Expectation value}
  \end{subfigure}
  \caption{\small Performance of the VQAs as well as RDGS for the maximization of the SNR $\rho$ as a function of the coordinates $(\theta_1,\eta)$. The mass parameter space $(\theta_1,\eta)$ is discretized using a uniform $256 \times 256$ grid. The results for the VQAs are averaged over $10$ repeats, with standard deviation indicated with error bars.}
  \label{fig:vqa_results}
\end{figure}

\begin{figure}[h]
  \centering
  \begin{subfigure}{0.47\linewidth}
    \centering
    \includegraphics[width=\linewidth]{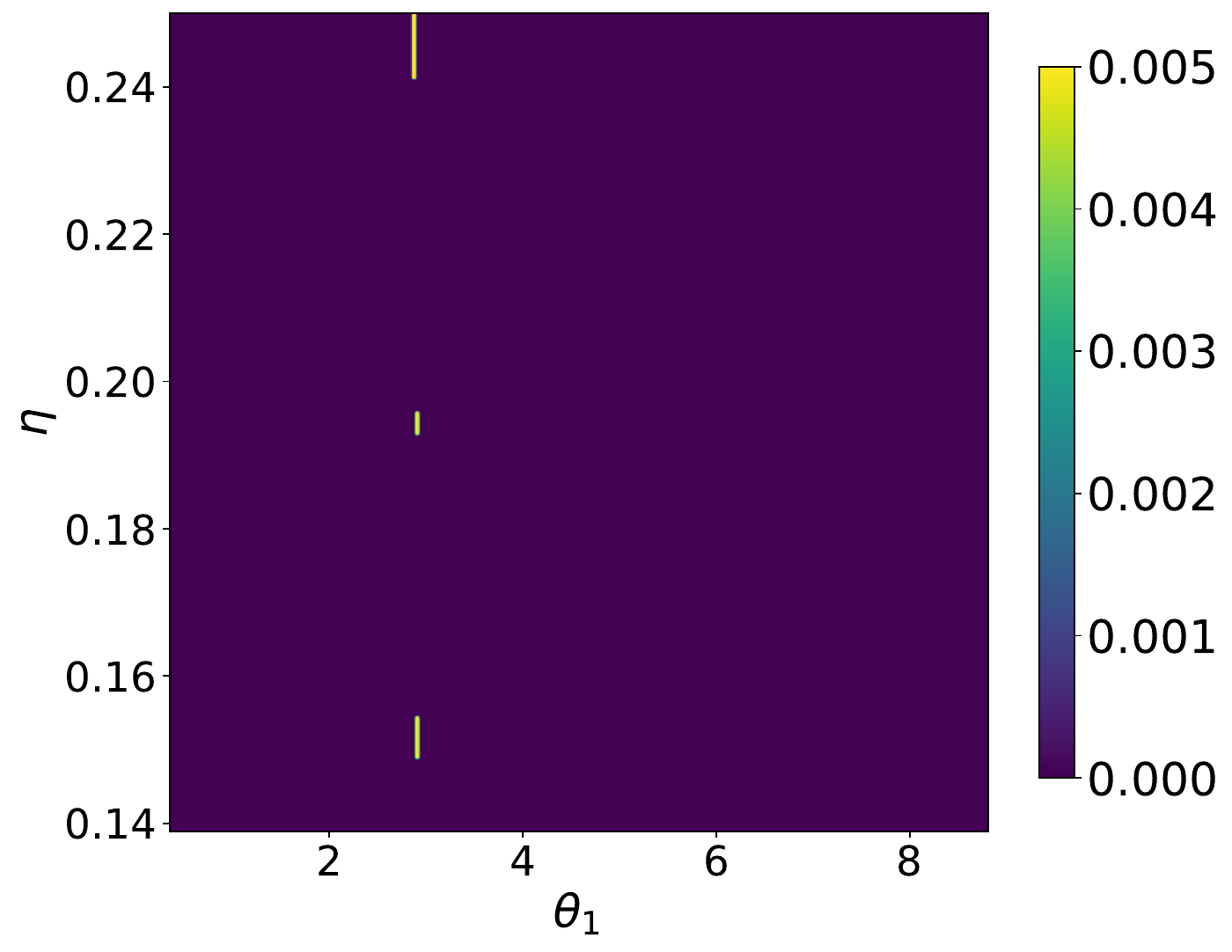}
    \caption{\small Final probabilities: RDGS}
  \end{subfigure}
  \hspace{5mm}
  \begin{subfigure}{0.47\linewidth}
    \centering
    \includegraphics[width=\linewidth]{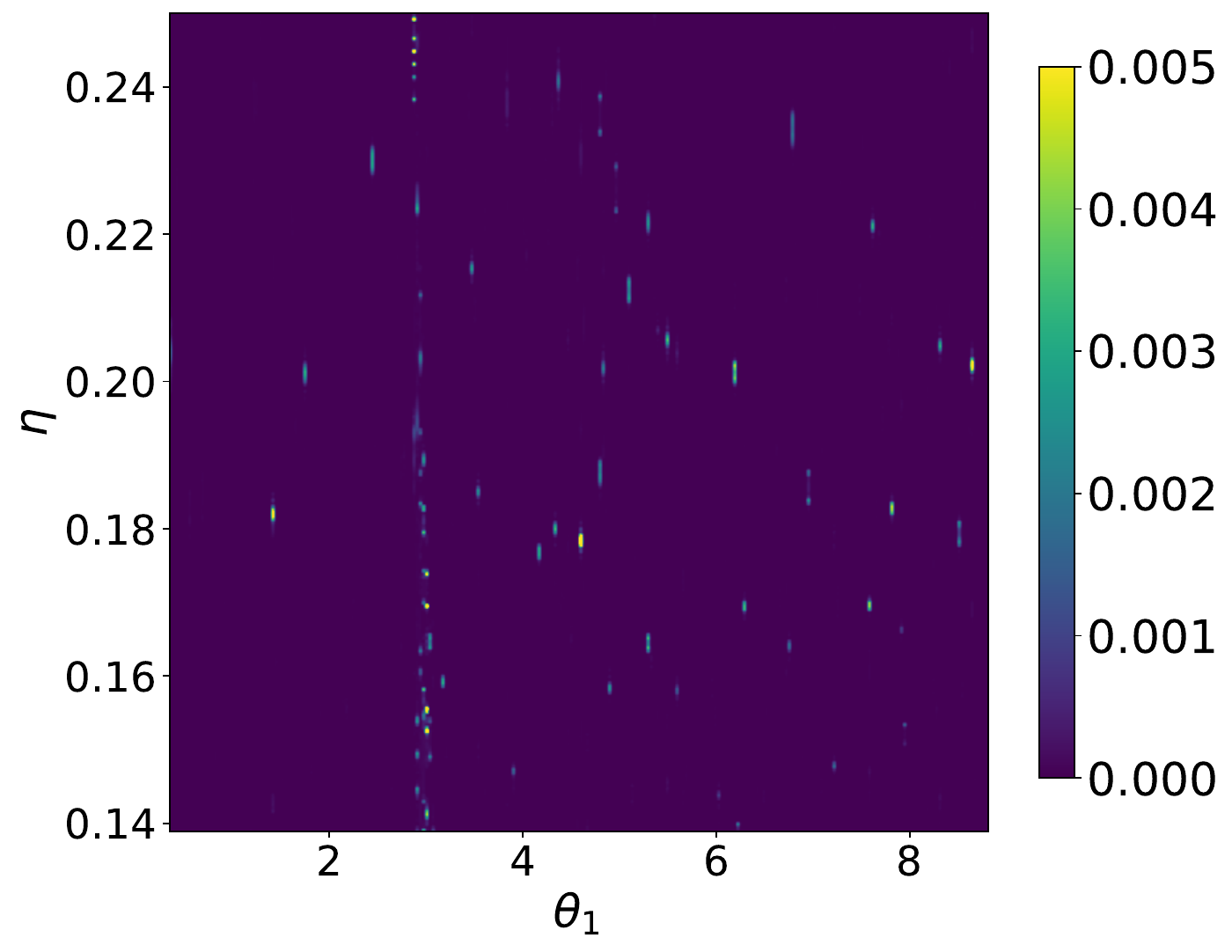}
    \caption{\small Final probabilities: QMOA (Complete)}
  \end{subfigure}
  \caption{\small Final probabilities (at depth $p=15$) for measuring mass parameters $(\theta_1,\eta)$. The mass parameter space $(\theta_1,\eta)$ is discretized using a uniform $256 \times 256$ grid. For QMOA (Complete), this is the state corresponding to the repeat which obtained the highest expectation value at $p=15$. The colorbar is scaled to easily see the states which are being amplified by QMOA (Complete).}
  \label{fig:vqa_finalprobs}
\end{figure}

The results of the simulation described above are presented in Figure~\ref{fig:vqa_results}.
Among the VQAs, the best performance, in terms of both probability of success and expectation value, is shown by QMOA (Complete) and QAOA (Complete).
However, all of the VQAs are significantly outperformed by the RDGS in terms of the probability of successfully identifying the GW signal in the time series.
Further, the difference in performance becomes more prominent with increasing depth.
Prior to a depth of $p = 8$, QMOA (Complete) and QAOA (Complete) arrive at an expectation value for the quality which is higher than that of RDGS, despite the probability of success being much lower.
This can be understood by plotting the final probabilities over the parameter space (see Figure~\ref{fig:vqa_finalprobs}).
In Figure~\ref{fig:vqa_finalprobs}, we see that RDGS only amplifies the states with a quality above the threshold, whereas QMOA (Complete) also amplifies the probabilities of many of the local maxima.
One can then interpret the results in Figure~\ref{fig:vqa_results} as demonstrating that QMOA (Complete) can easily find the local maxima for small depth ($p \leq 8$), but fails to significantly transfer probability from these local maxima to the states with a quality above the threshold.

\begin{figure}[h]
  \centering
  \includegraphics[width=0.5\linewidth]{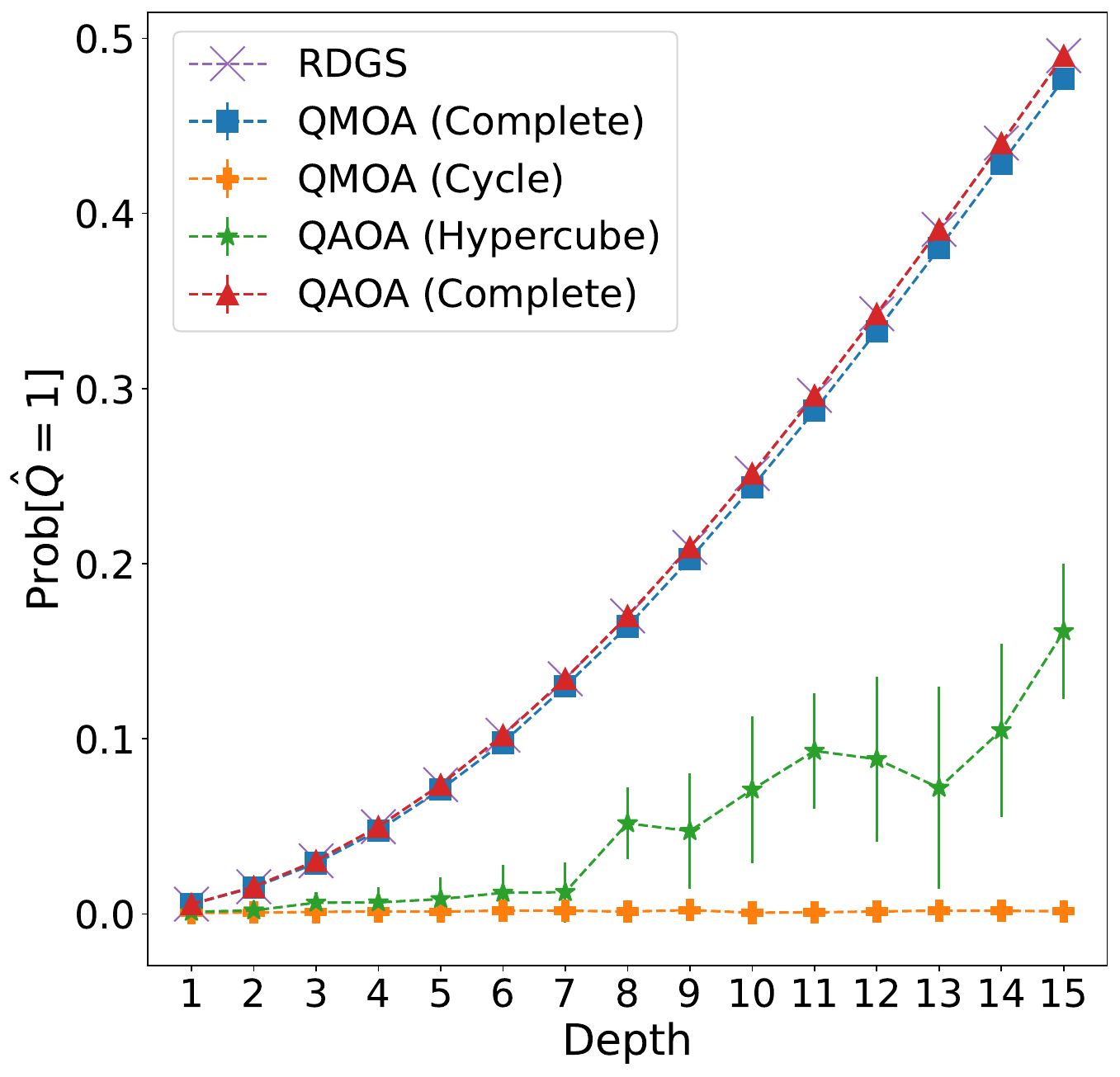}
  \caption{\small Performance of the VQAs as well as RDGS for the maximization of a binary cost function over the coordinates $(\theta_1,\eta)$. The mass parameter space $(\theta_1,\eta)$ is discretized using a uniform $256 \times 256$ grid. The results for the VQAs are averaged over $10$ repeats, with standard deviation indicated with error bars.}
  \label{fig:vqa_results_binary}
\end{figure}

We also completed simulations where the cost function of the VQAs was replaced with the binary cost function used in Grover's algorithm.
That is, we replace
\begin{equation}
  \hat{Q} = \sum_{\theta_1,\eta} \rho(\theta_1,\eta) \ket{\theta_1,\eta}\bra{\theta_1,\eta}
\end{equation}
with
\begin{equation}
  \hat{Q} = \sum_{\{ \theta_1,\eta \text{ $:$ } \rho(\theta_1,\eta) > \rho_0 \}} \ket{\theta_1,\eta}\bra{\theta_1,\eta}.
\end{equation}
This is not ideal for a VQA, since estimating the expectation value requires taking a large number of samples.
However, we were interested in examining whether the VQAs could improve upon RDGS if the local maxima in the quality function are removed.
Also, in this case, the variational optimization aims to directly maximize the probability of successfully identifying a signal, since it coincides with the expectation value.
As shown in Figure~\ref{fig:vqa_results_binary}, with a binary cost function, QMOA (Complete) and QAOA (Complete) were able to match the performance of Grover's algorithm, but not exceed it.

\begin{figure}[h]
  \centering
  \begin{subfigure}{0.45\linewidth}
    \centering
    \includegraphics[width=\linewidth]{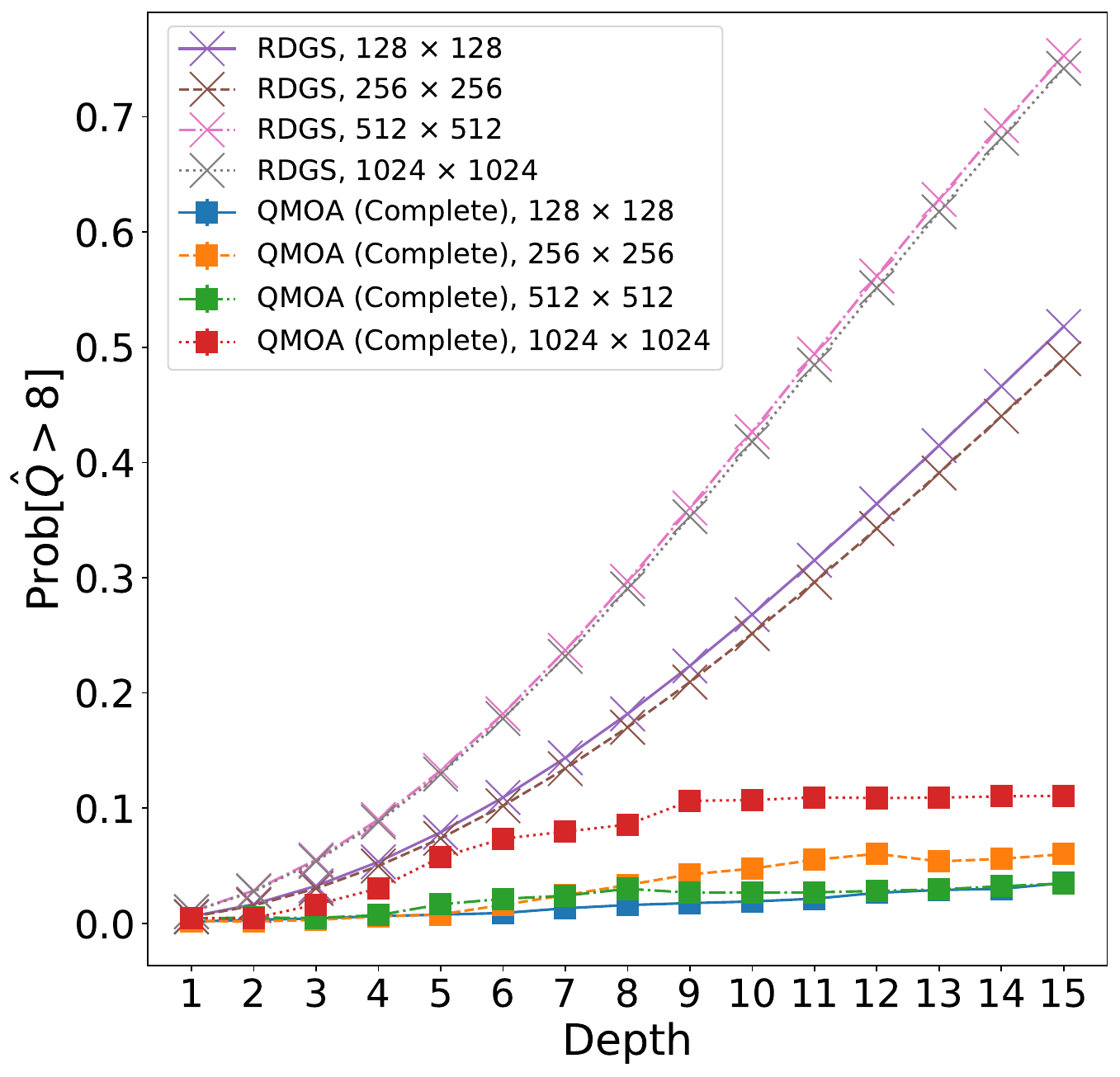}
    \caption{\small Probability of success}
  \end{subfigure}
  \hspace{5mm}
  \begin{subfigure}{0.45\linewidth}
    \centering
    \includegraphics[width=0.95\linewidth]{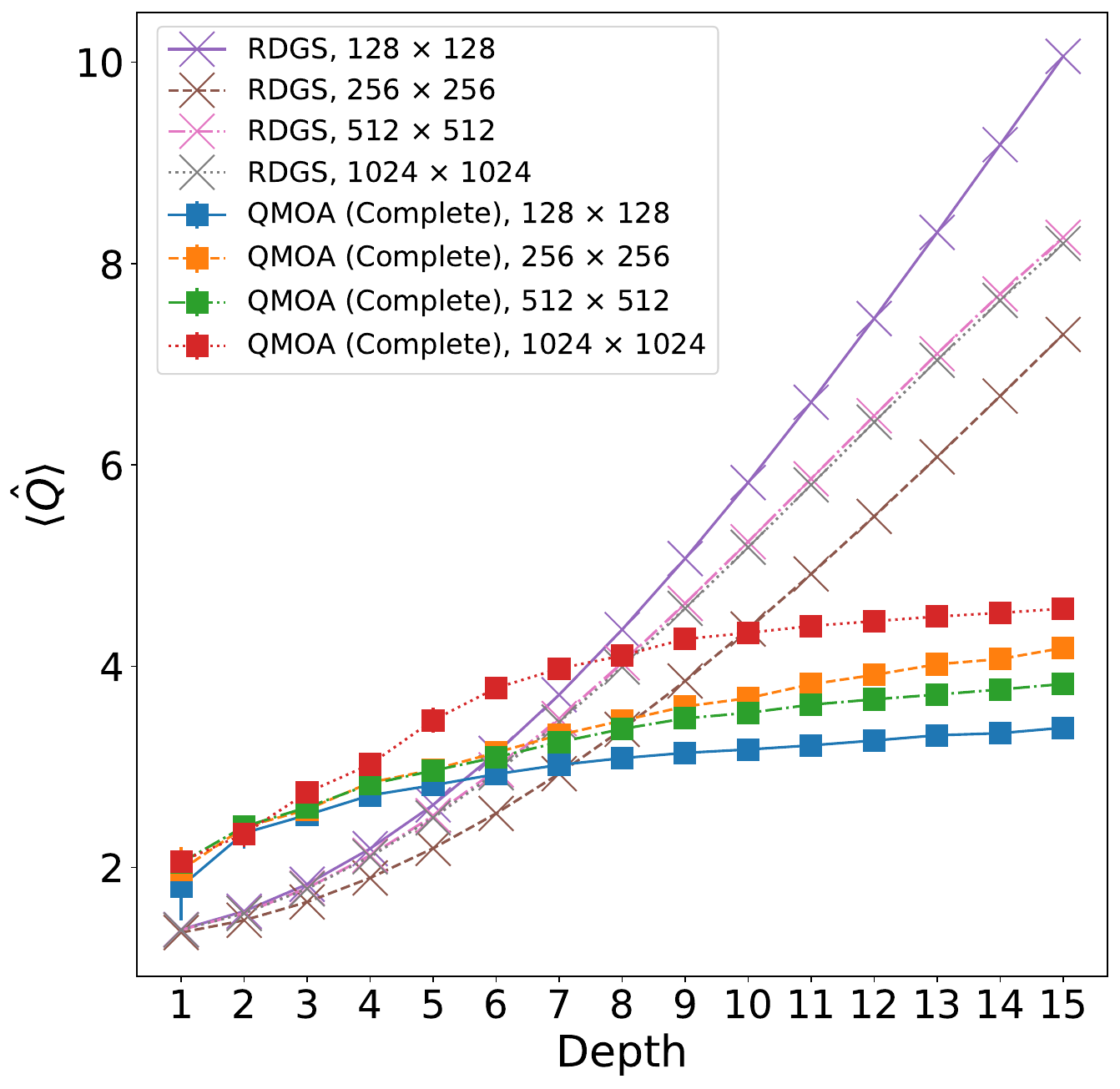}
    \caption{\small Expectation value}
  \end{subfigure}
  \caption{\small Performance of QMOA (Complete) RDGS for the maximization of the SNR $\rho$ as a function of the coordinates $(\theta_1,\eta)$, with grid size ranging from $128 \times 128$ (14 qubits) to $1024 \times 1024$ (20 qubits). The results for QMOA (Complete) are averaged over $10$ repeats, with standard deviation indicated with error bars.}
  \label{fig:resolution}
\end{figure}

We also considered the possibility that the VQAs are not able to exploit structure in the parameter space because the discretization is too coarse.
We therefore examined the performance of QMOA (Complete), which was the best-performing of the structured VQAs, as the grid resolution is increased from $128 \times 128$ (14 qubits) to $1024 \times 1024$ (20 qubits).
However, Figure~\ref{fig:resolution} demonstrates that the results do not change significantly.

\begin{figure}[h!]
  \centering
  \begin{subfigure}{0.45\linewidth}
    \centering
    \includegraphics[width=\linewidth]{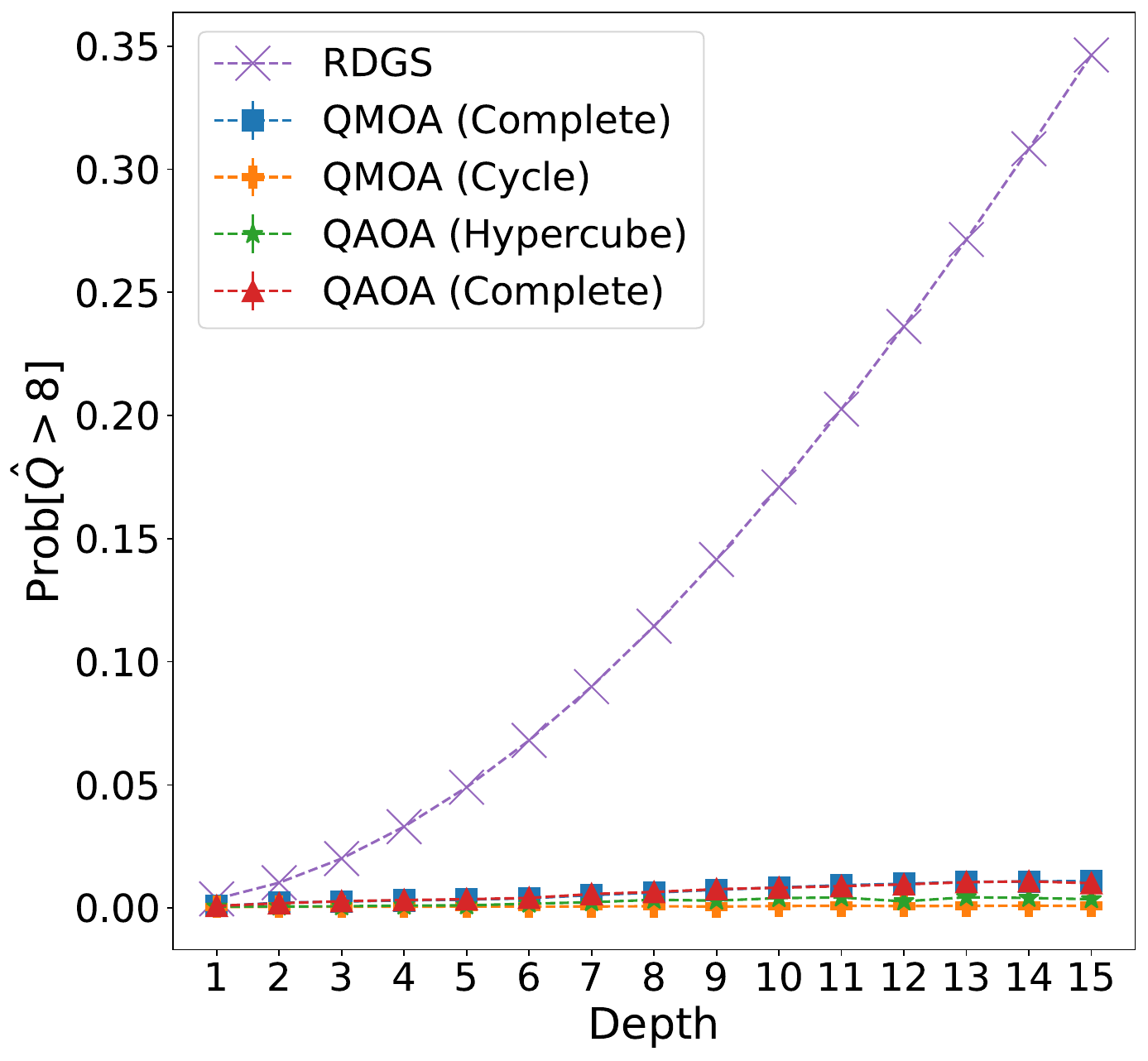}
    \caption{\small Probability of success}
  \end{subfigure}
  \hspace{5mm}
  \begin{subfigure}{0.45\linewidth}
    \centering
    \includegraphics[width=0.97\linewidth]{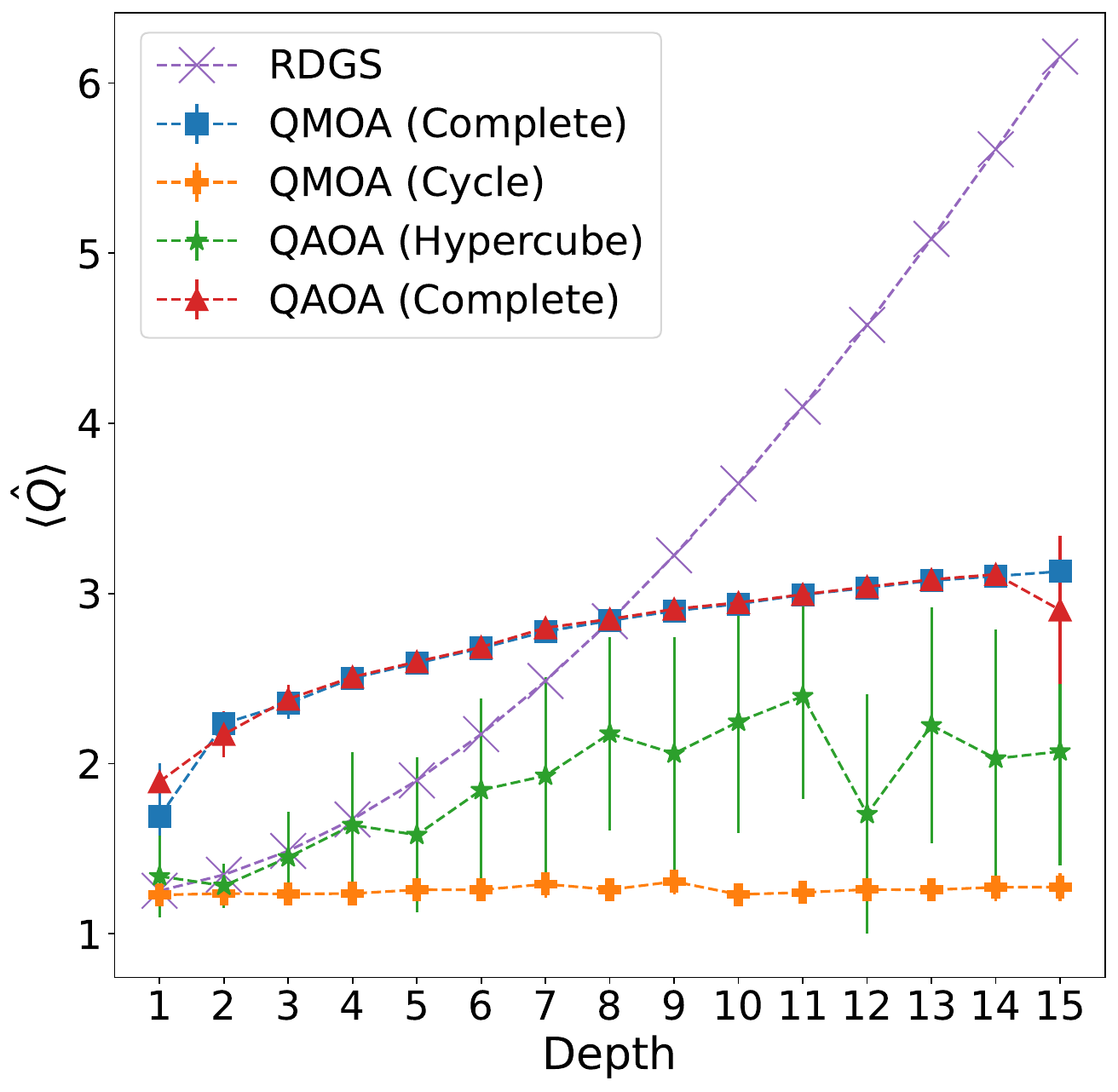}
    \caption{\small Expectation value}
  \end{subfigure}
  \caption{\small Performance of the VQAs as well as RDGS for the maximization of the SNR $\rho$ as a function of the coordinates $(m_1,m_2)$. The mass parameter space $(m_1,m_2)$ is discretized using a uniform $256 \times 256$ grid. The results for the VQAs are averaged over $10$ repeats, with standard deviation indicated with error bars.}
  \label{fig:vqa_results_theta1eta}
\end{figure}

Finally, we examined the VQA performance in different mass coordinates, to see if the VQAs would be able to capture structure in the cost function with a different representation.
To this end, we reexpressed the quality function $\rho$ in terms of the $(m_1,m_2)$ coordinates, and simply employed a uniform discretization of these parameters with a $256 \times 256$ grid in the region $1 \, M_\odot \leq m_1, m_2 \leq 5 \, M_\odot$.
Hence, the quality function is that shown in Figure~\ref{fig:GW170817_snr_m1m2}.
The VQA performance results are shown in Figure~\ref{fig:vqa_results_theta1eta}.
We see that the performance of the VQAs, as compared to RDGS, does not change significantly.
We also completed simulations with $(M,\eta)$ and $(\theta_1,\theta_2)$ coordinates (where $\theta_2 := \tfrac{\pi}{4} (\pi G M f_s / c^3)^{-2/3} \eta^{-1}$ \cite{OwenSathyaprakash1999}), but the results were similar.

\section{Discussion}
\label{sec:discussion}

It appears from the results in the previous section that the structured VQA approaches do not outperform the RDGS.
Is this because they are not capturing the structure present in the search problem, or is it simply because there is no structure to be captured?

\begin{figure}[h]
  \centering
  \begin{subfigure}{0.47\linewidth}
    \centering
    \includegraphics[width=\linewidth]{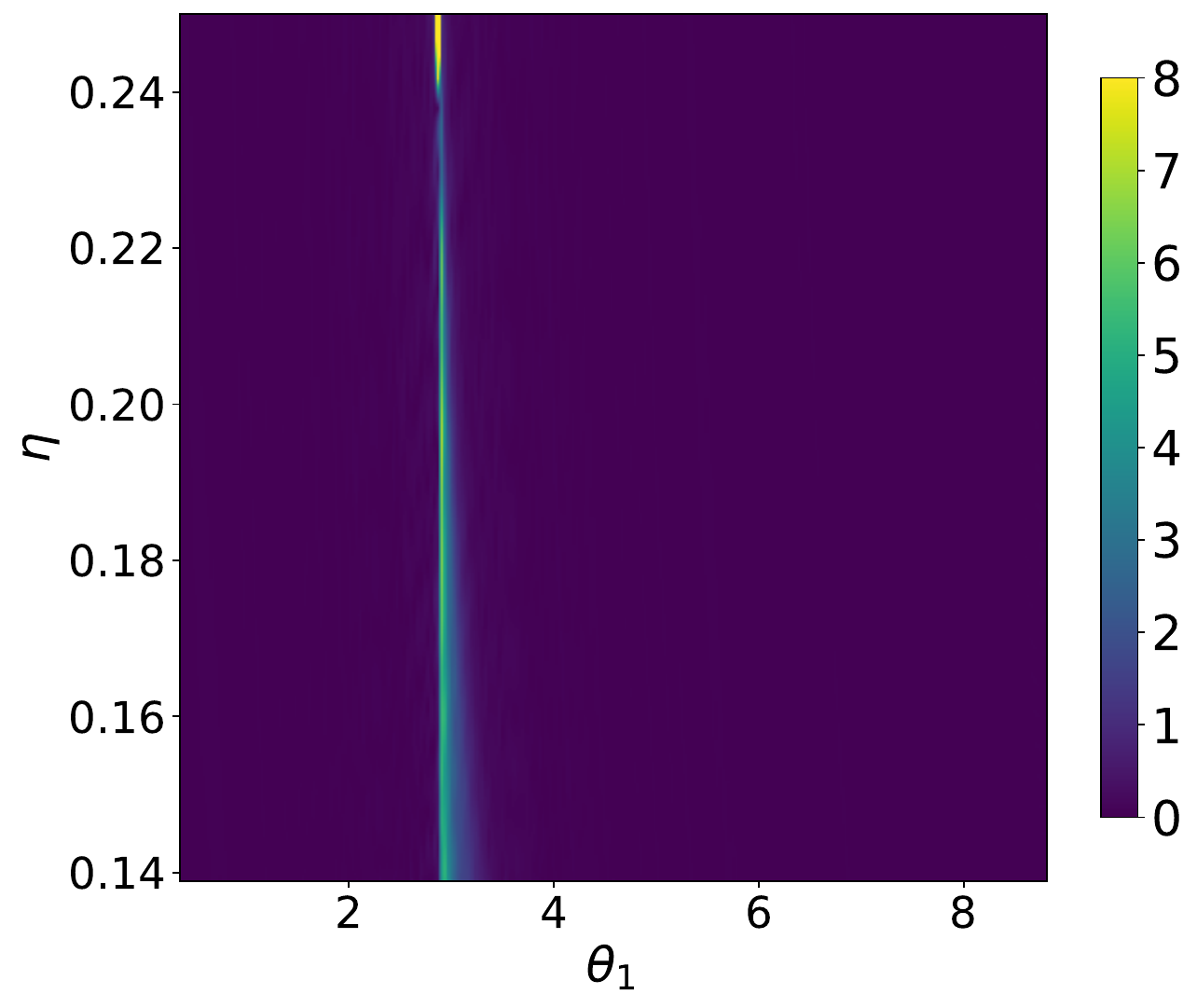}
    \caption{\small Signal only}
  \end{subfigure}
  \hspace{5mm}
  \begin{subfigure}{0.47\linewidth}
    \centering
    \includegraphics[width=\linewidth]{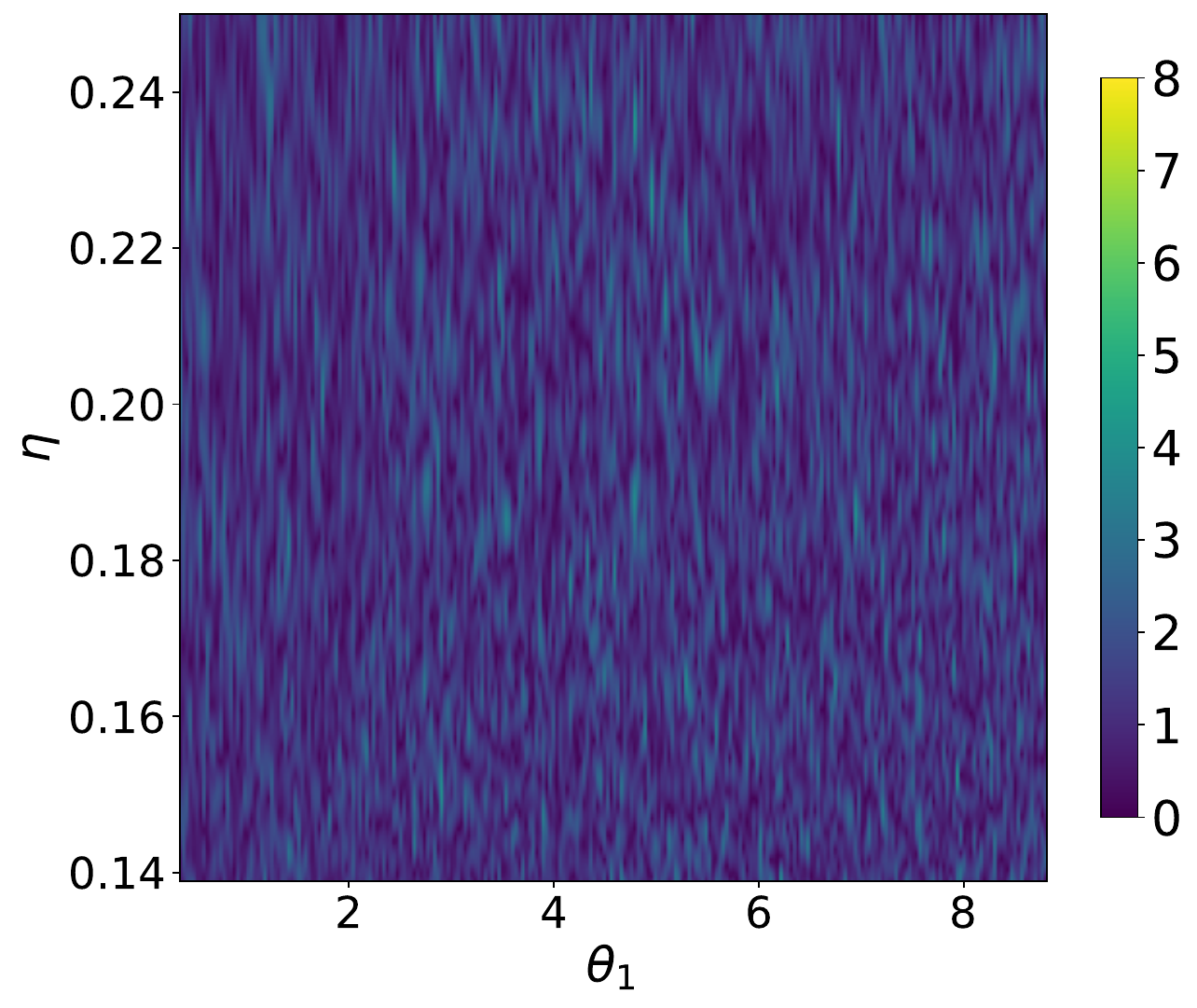}
    \caption{\small Noise only}
  \end{subfigure}
  \caption{\small Decomposition of the GW170817 quality function (Figure~\ref{fig:GW170817_snr_theta1eta}) into signal and noise in $(\theta_1,\eta)$ coordinates. (a) The quality function for a time series of a noiseless signal: $y \to (h|y) h$. (b) The quality function for the time series with the signal subtracted: $y \to y - (h|y) h$.}
  \label{fig:signal_noise_decomp_GW170817}
\end{figure}

To get some insight, one can examine the difference in the quality function with and without a signal present (see Figure~\ref{fig:signal_noise_decomp_GW170817}).
To analyze this, we evaluated the quality function $\rho$ over the parameters $(\theta_1,\eta)$ with the time series $y$ replaced with its components along and orthogonal to the template $h$ corresponding to the parameters of the signal $m_1 = m_2 = 1.3758 \, M_\odot$ (i.e., $\theta_1 \approx 2.87$ and $\eta = 0.25$).
That is, we replaced the time series with one associated with a noiseless signal, $y \to (h|y) h$, and one associated with purely noise, $y \to y - (h|y) h$.
It is clear from Figure~\ref{fig:signal_noise_decomp_GW170817} that most of the variation in the quality values across the parameter space is due to noise in the time series, and is present whether or not there is a signal.
One can also see that the presence of a signal is only indicated along a relatively small band of parameter values (which corresponds to a line of constant chirp mass).
Therefore, it seems there is very little structure present in the quality function which can be exploited in order to guide a search.
This is consistent with our finding that the RDGS performs better than the structured VQA searches for this problem.

\begin{figure}[h]
  \centering
  \begin{subfigure}{0.47\linewidth}
    \centering
    \includegraphics[width=\linewidth]{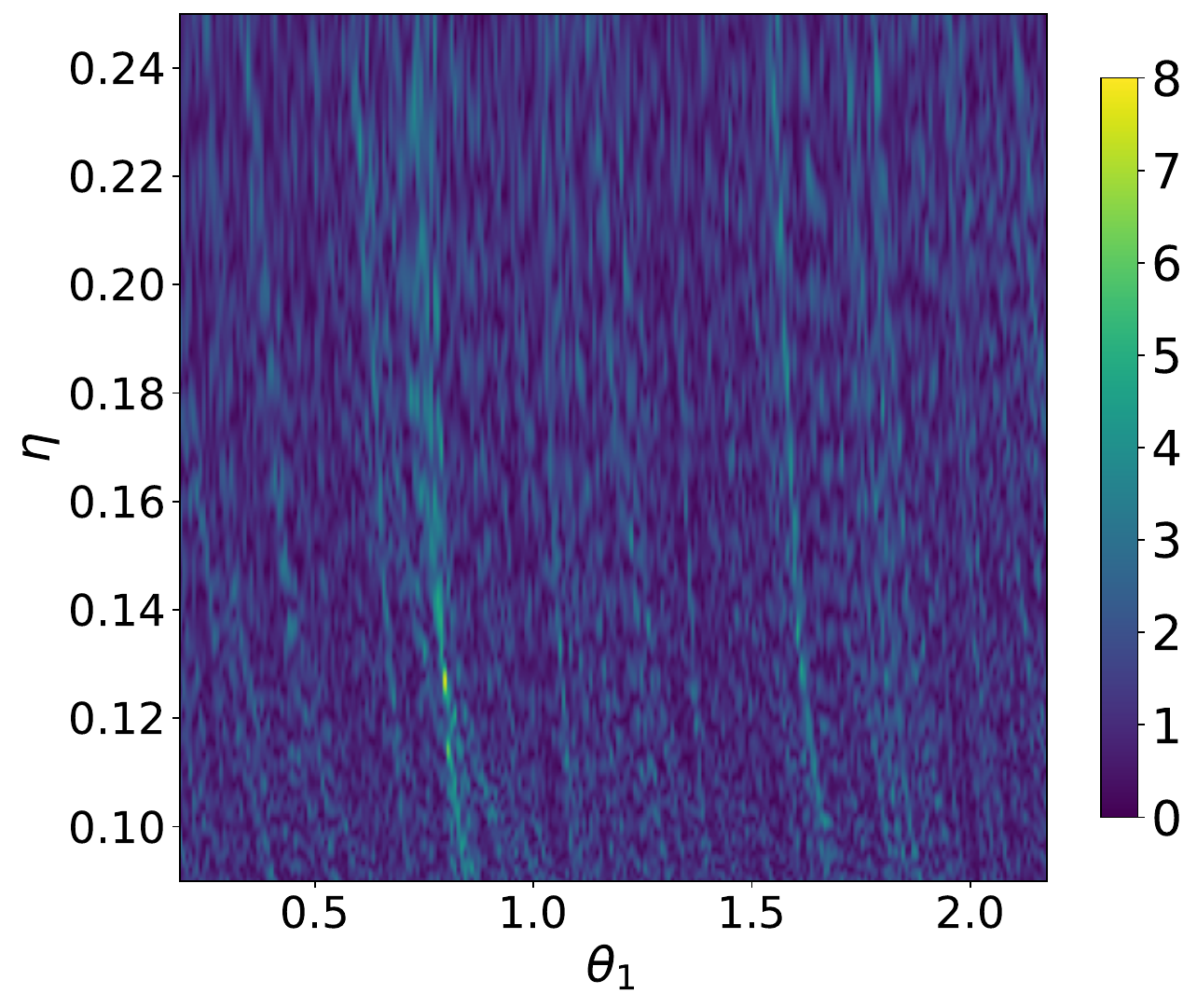}
    \caption{\small Quality function}
  \end{subfigure}
  
  \begin{subfigure}{0.47\linewidth}
    \centering
    \includegraphics[width=\linewidth]{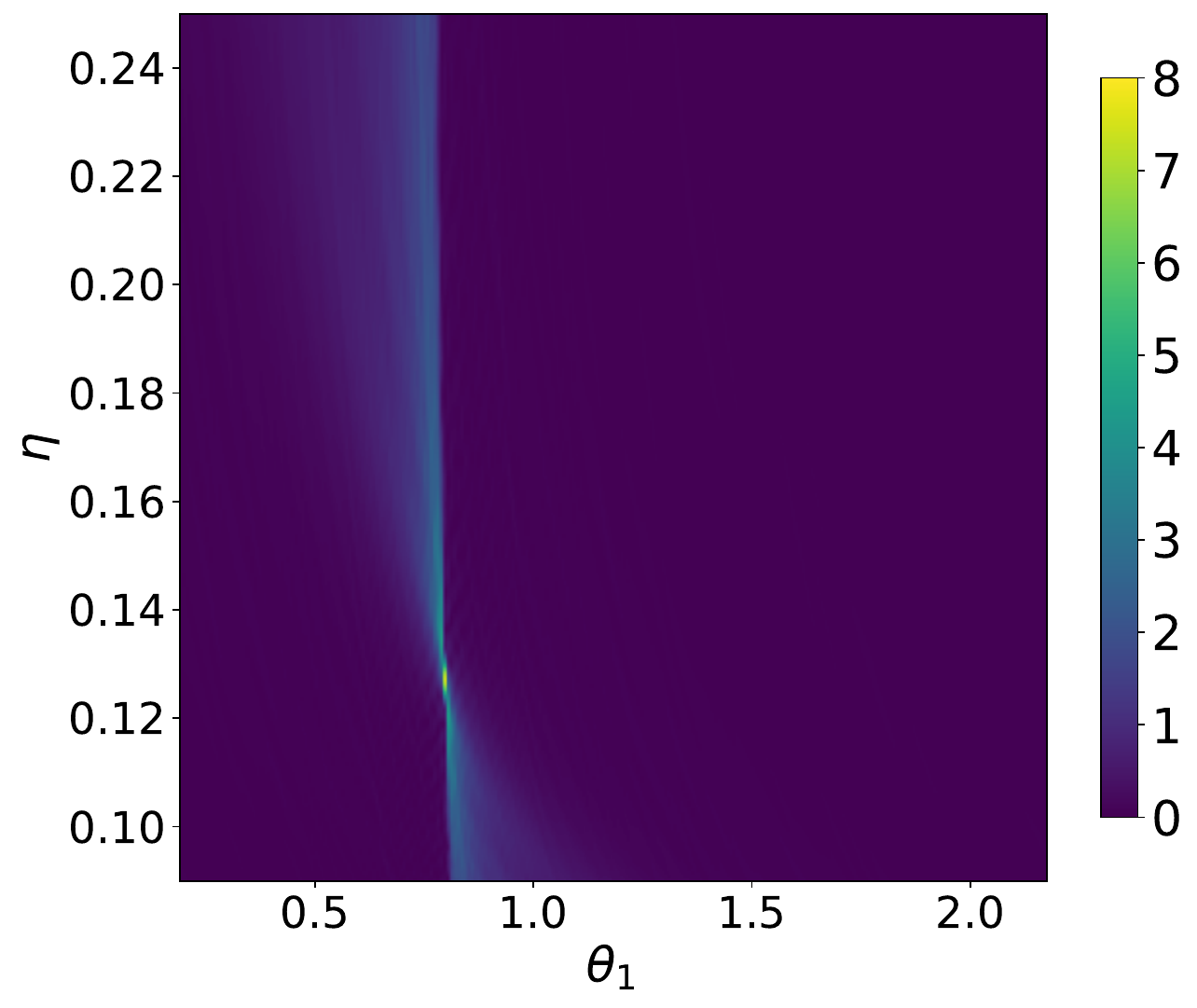}
    \caption{\small Signal only}
  \end{subfigure}
  \hspace{5mm}
  \begin{subfigure}{0.47\linewidth}
    \centering
    \includegraphics[width=\linewidth]{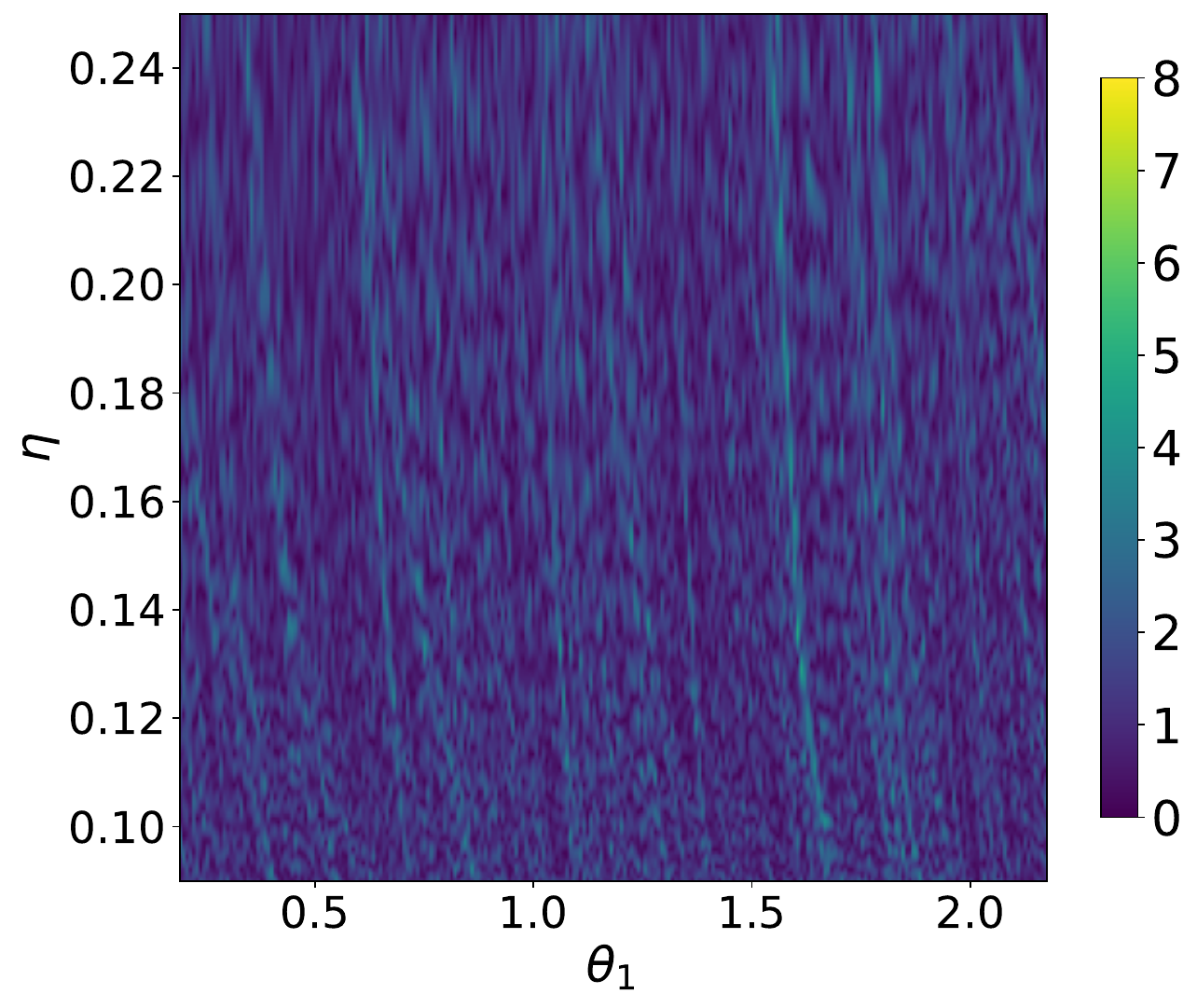}
    \caption{\small Noise only}
  \end{subfigure}
  \caption{\small Decomposition of the GW200115\_042309 quality function into signal and noise in $(\theta_1,\eta)$ coordinates. (a) The quality function for a time series $y$ containing the GW event. The parameters corresponding to the maximum of the quality function are: $\theta_1 \approx 0.797$ and $\eta \approx 0.127$ (corresponding to $m_1 \approx 7.58 \, M_\odot$ and $m_2 \approx 1.33 \, M_\odot$). (b) The quality function for a time series of a noiseless signal: $y \to (h|y) h$. (c) The quality function for the time series with the signal subtracted: $y \to y - (h|y) h$.}
  \label{fig:signal_noise_decomp_GW200115_042309}
\end{figure}

We also note that the event GW170817 has a relatively large SNR, compared to other GW events.
A consequence of this is that there are multiple parameter values which yield a quality value above the threshold.
For events with smaller SNR, such as the event GW200115\_042309 \cite{AbbottEtal2023} (see Figure~\ref{fig:signal_noise_decomp_GW200115_042309}), the resulting quality function resembles even more closely an unstructured search problem, since one has to be in close proximity to the signal parameters in the space in order to determine whether there is a signal.
In other words, features in Figure~\ref{fig:signal_noise_decomp_GW200115_042309}(b) which may help to guide a search towards the maximum are not apparent when combined (Figure~\ref{fig:signal_noise_decomp_GW200115_042309}(a)) with contributions of the noise from Figure~\ref{fig:signal_noise_decomp_GW200115_042309}(c).

\section{Conclusion}
\label{sec:conclusion}

In this paper, we explored the potential of variational quantum algorithms (VQAs) in speeding up the template search problem which arises in the detection of gravitational wave signals.
Despite previously demonstrated successes of VQAs for certain classes of optimization problems, here we found that a simple restricted-depth Grover search exceeded the performance of the tested VQAs.
It appears that the GW template search, as we implemented it, essentially constitutes an unstructured search problem (i.e., lacking exploitable structure). Therefore, the use of a restricted-depth Grover search (RDGS) algorithm seems optimal for this task.
Although the VQAs were not able to improve upon the performance of the RDGS for this problem, in the search for applications of VQAs as well as near-term quantum devices, we believe it is important to also understand the limitations of VQAs.

\section*{Acknowledgments}

JP was supported by the Brian Dunlop Physics Fellowship at The University of Western Australia and is supported at Nordita by the Wenner-Gren Foundations and, in part, by the Wallenberg Initiative on Networks and Quantum Information (WINQ).
Nordita is supported in part by NordForsk.
MK was supported by funding from the Australian Research Council (ARC) Centre of Excellence for Gravitational Wave Discovery (OzGrav).
This work is supported by resources provided by the Pawsey Supercomputing Research Centre with funding from the Australian Government and the Government of Western Australia. The authors would also like to thank Tavis Bennett for helpful discussions.

\bibliography{main}

\end{document}